\documentclass[aps,pre,twocolumn]{revtex4-1}

\usepackage{graphicx}
\usepackage{dcolumn}
\usepackage{bm}

\usepackage[utf8]{inputenc}
\usepackage{mathtools}
\usepackage[T1]{fontenc}
\usepackage{mathptmx}
\usepackage[utf8]{inputenc}
\usepackage{graphicx}
\usepackage{subcaption}
\usepackage{amsmath}
\usepackage{amssymb}
\usepackage{dcolumn}
\usepackage{bm}
\usepackage{xr}
\usepackage{cleveref}

\begin{document}

\title{Low-entangled UHMWPE melt: analytical model and computer simulations}
\author{Artem Petrov}
\email{petrov.ai15@physics.msu.ru}
	\affiliation{Faculty of Physics, Lomonosov Moscow State University, 119991 Moscow, Russia}
	\author{Pavel Kos}
	\affiliation{Faculty of Physics, Lomonosov Moscow State University, 119991 Moscow, Russia}
	\affiliation{Semenov Institute of Chemical Physics, 119991 Moscow, Russia}
	\author{Vladimir Rudyak}
	\affiliation{Faculty of Physics, Lomonosov Moscow State University, 119991 Moscow, Russia}
	\author{Alexander Chertovich}
	\affiliation{Semenov Institute of Chemical Physics, 119991 Moscow, Russia}
	\affiliation{Faculty of Physics, Lomonosov Moscow State University, 119991 Moscow, Russia}
	\date{\today}

\begin{abstract}
In this work we developed a theoretical model to describe the polymerization of very long macromolecules with simultaneous precipitation. As a reference system for our model we consider UHMWPE polymerization process, via homogeneous catalysis and with molecular weight above $10^6$ monomers. We derived time dependency of the entanglement length in the system, and assessed how polymerization rate and fraction of initiators affects this dependency. We show that decrease of the fraction of initiators leads to increase of the average entanglement length in the melt after complete polymerization. The results are supported by computer simulations and in principle could be applied to the wide set of polymerized materials, both crystallized and amorphous.
\end{abstract}

\maketitle

\section{Introduction}
One of the unclear topics in the modern polymer physics is the question how the polymerization process affects the structure of the resulting polymer material. This question is also connected to the subject of polymer sample history and its influence on the rheological and mechanical properties. The answer is clear for short enough molecules, which resemble either good solvent or poor solvent conformations in solution and very quickly arrive to gaussian statistics in a melt, according to a Flory theorem. But this is not the case for long enough macromolecules, where the entanglements between polymer segments play important role and the time scaling to reach thermodynamic equilibrium is a power law from the molecular length N: $t\propto N^{3.4}$. Thus, the equilibration time for really long macromolecules run out from the observation time and there is no way to anticipate equilibrium statistics for various systems from long macromolecules by default. 

The characteristic polymer length, from which the aforementioned non-equilibrium and history-dependent properties become obvious, are usually starts from $N\approx10^5$ monomer units, which is large enough and up to now most typical laboratory and industrial macromolecules lies below this threshold. But there is one important exception: Ultra High Molecular Weight PolyEthylene (UHMWPE), also known as high-modulus polyethylene (HMPE), it has extremely long chains, with a molecular mass usually above 5 million Daltons (i.e. more than $3\times10^5$ monomer units). This unique material has a great strength (the strength-to-weight ratios of it fibers is around eight times that of high-strength steels), perfect chemical and abrasion resistant, and already produced nowadays in various forms by the hi-tech industry. The only problem is that the typical UHMWPE fiber production technology, initially developed by Leemstra and Smith \cite{smith1980ultra} and commercialized by DSM company, is the complex and environment unfriendly solution spinning process, which resulted also in a great cost. Solution spinning process plays on isolating individual chain molecules in the solvent by dilution to minimize intermolecular entanglements, which make chain orientation more easy, and increase the strength of the final product.

The first paper where the direct synthesis of UHMWPE with already a reduced number of entanglements has been described by Smith et al. \cite{smith1985drawing}. The authors observed that the drawing properties of the material depend on the polymerization temperature: samples synthesized at lower temperatures show an increased drawability and they attributed this to the reduced number of entanglements. This study proved that directly synthesized low-entangled UHMWPE can be processed below the melting temperature of the polymer to obtain fibers and tapes with very high draw ratio, high modulus and high strength without the use of any solvent.

The next step was done more recently by Rastogi and coworkers using a homogeneous single-site catalytic systems \cite{rastogi2011unprecedented}. It is based on post-metallocene catalysts, which is soluble in reaction media and together with co-catalyst can give 10 – 100 times higher activity than the classical Ziegler–Natta systems. It was shown, that in some cases the polymerization process could be considered as a living process, without termination and chain transfer events \cite{talebi2008disentangled}. Moreover, it was directly shown that the decrease of catalyst concentration (i.e. number of growing chains per unit volume) increase both the molecular weight and yield stress \cite{pandey2011heterogeneity}. The main concept is that, when using a single-site catalyst instead of a heterogeneous one, it is possible to control the proximity of the active sites and the reaction rate by changing the reaction conditions (catalyst concentration, monomer pressure, polymerization time, temperature). 

Summarizing, tailoring reaction conditions (i.e. temperature, pressure, reaction time, amount of solvent, type of catalyst, concentration) it is possible to tune the characteristics of the polymer synthesized, including the entanglement density. The main goal could be formation of low-entangled polymer melt directly in the reaction media, following by the solid-state deformation to prepare UHMWPE fibers in a cheap, reproducible and environmentally friendly way. Nevertheless, there is still no clear relationship between reaction conditions and properties of the reaction product. Most work done in many experimental attempts to select more suitable catalyst system, but without any suitable model of the system under study. An additional drawback of Rastogi et al. approach \cite{rastogi2011unprecedented} is that the final product is still a powder with great abundance of a solvent in a reaction media. Thus there is a great dependence on the powder processing before fiber drawing (i.e. solvent evaporation scheme, compression intensity/temperature).

We can mention here probably the only one theoretical model to describe formation of non-nequilibrium and low-entangled polymer melt \cite{mcleish2007theory}. In this work author describe the kinetic trapping during the aggregation of single chain crystallites: initially crumpled conformations want to spread upon melt formation, similar to osmotic explosion process initially proposed by De Gennes, but this spreading stops very quickly as soon as a narrow region of melt with Gaussian statistics and corresponding entanglements is formed.

We present here the model to describe the process of UHMWPE precipitation polymerization via homogeneous catalysis and the approximation of living chains growth. With the help of the model it is possible to predict what should be the polymerization degree before the intermolecular entanglements starts to appear and what should be the overall entanglement length of the system if we consider growth of extremely long macromolecules until the united melt is formed.

\section{Theoretical model}
Let us consider the radical homopolymerization process taking place in a poor solvent. In the beginning of the process, the system consists of the active sites, monomers and solvent. The fraction of active sites (or fraction of initiators) $c$ is defined as the number of polymerization initiators divided by the number of particles in the system. For simplicity, we assume that all active sites are initiated simultaneously, and there is no spontaneous termination of the growth as well as no chain transfer. Thus $c$ is constant over the reaction time $t$. The reaction begins, and the chains start to grow, and the average chain length $N$ increases. We will define the reaction rate $r$ as the average number of monomers, attached to a growing end of a chain in a time unit: $r = dN/dt$. We consider the reaction rate $r$ is constant over time, so $N = rt$ (Fig. \ref{conversion}). We assume, that there are two distinct phases during the growing process: 1) chains grow without meeting each other, and, thus, forming no entanglements ($N_e = N$); 2) when some critical value of the average chain length $N_c$ is reached, the chains start to intersect and form entanglements.

This section is organized as follows: in the subsections II.A and II.B we make assumptions regarding the first phase of the chain growth, when each chain can be treated as an isolated growing chain in a poor solvent. Basing on our assumptions, we develop the analytical model, which describes the dependency of $N_c$, the average chain length when chains start to form entanglements, on the reaction rate $r$ and the fraction of active sites $c$. Then in the subsection II.C we develop a simple model of how entanglements are formed after chains met each other (i.e. in the second phase), and how $N_c(r,c)$ defines the average entanglement length in the system. In the subsection II.D we build the unified $N_c(r,c)$ dependency, which is compared with computer simulations results, and which defines the dependency of the entanglement length after polymerization on $r$ and $c$ ($N_e|_{N=1/c}(r,c)$) (see Computer Simulation Results section).

\subsection{Large fraction of initiators}
It was established, that there is a strong molecular weight dependence of the coil-globule transition temperature for PNIPAM (poly(N-isopropylacrylamide)) \cite{xia2005thermal,shan2009thermoresponsive}. The transition temperature starts to sharply increase for oligomers length $\propto 10$ monomers \cite{xia2005thermal,shan2009thermoresponsive}. Since PNIPAM chains are considered flexible \cite{xia2005thermal,shan2009thermoresponsive}, we can roughly say, that molecular weight of the polymer starts to affect the collapse temperature of chains length $N<10$ Kuhn segments. In our work we are dealing with growing chains in a poor solvent below the crystallization temperature, so the Kuhn segment length increases up to $\approx 10$ beads for non-collapsed chains (Fig. \ref{coilglob}). Therefore, if the growing chain is shorter than $N_{min}\approx 100$ beads, its collapse temperature will be shifted. If the temperature in the system is slightly below the coil-globule transition temperature for long chains (see Supporting Information for details), we can assume, that for the chains length $N<N_{min}$ the transition temperature will be shifted strong enough to hamper their collapse. In this approximation we can say, that the conformation of the chain length $N<N_{min}$ in our system is Gaussian with persistence length $n_p\approx10$ beads, and the end-to-end distance in such conformation is $R=an_p(N/n_p)^{1/2}=a(Nn_p)^{1/2}$. In our simple model, we do not take into account the diffusion of the chains as a whole. We assume, that interchain contact happens, if spheres with radius $R(N_c)$ become densely packed (i.e. when chains start to overlap). In other words, $cR^3(N_c)\propto1$, and by definition of $N_c$,  $c(an_p(N_c/n_p)^{1/2})^3\propto1$. Therefore, we obtain Equation \ref{eq:1}.

\begin{equation}
\label{eq:1}
    \begin{cases}
    N_c(c)\propto c^{-2/3}\\
    N_c(c)<N_{min}\approx 100
    \end{cases}
\end{equation}

This formula with exact coefficients is given in the Application 1. Since the overlap concentration governs the formation of entanglements in dilute solutions \cite{milner2005predicting}, we assume that the chains longer than $N_c$ become entangled with each other.

\subsection{Small fraction of initiators}
\begin{figure}[h!]
    \centering
	\begin{subfigure}{0.2\textwidth}
	\includegraphics[width=\linewidth,height=\textheight,keepaspectratio]{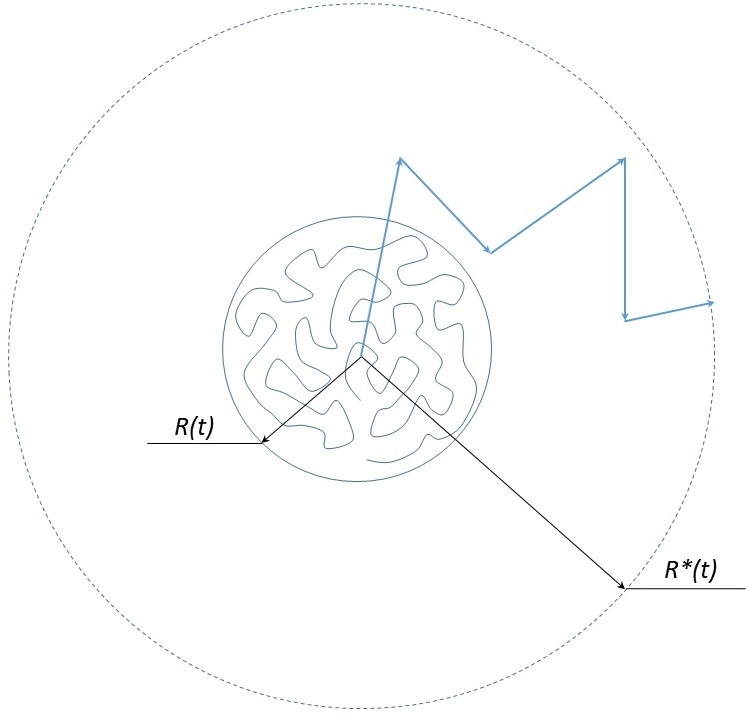}
	\end{subfigure}
	\begin{subfigure}{0.2\textwidth}
	\includegraphics[width=\linewidth,height=\textheight,keepaspectratio]{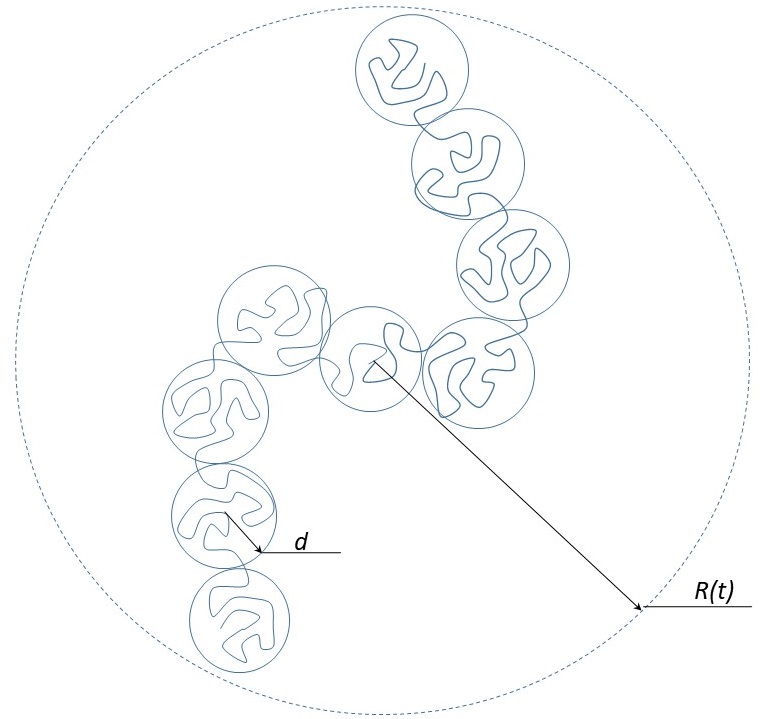}
	\end{subfigure}
	\begin{subfigure}{0.2\textwidth}
	\includegraphics[width=\linewidth,height=\textheight,keepaspectratio]{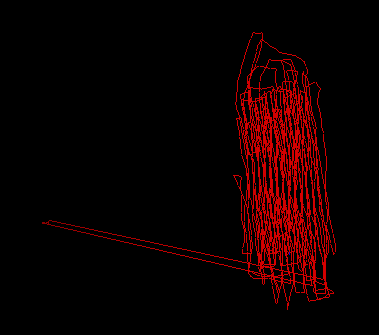}
	\end{subfigure}
	\begin{subfigure}{0.2\textwidth}
	\includegraphics[width=\linewidth,height=\textheight,keepaspectratio]{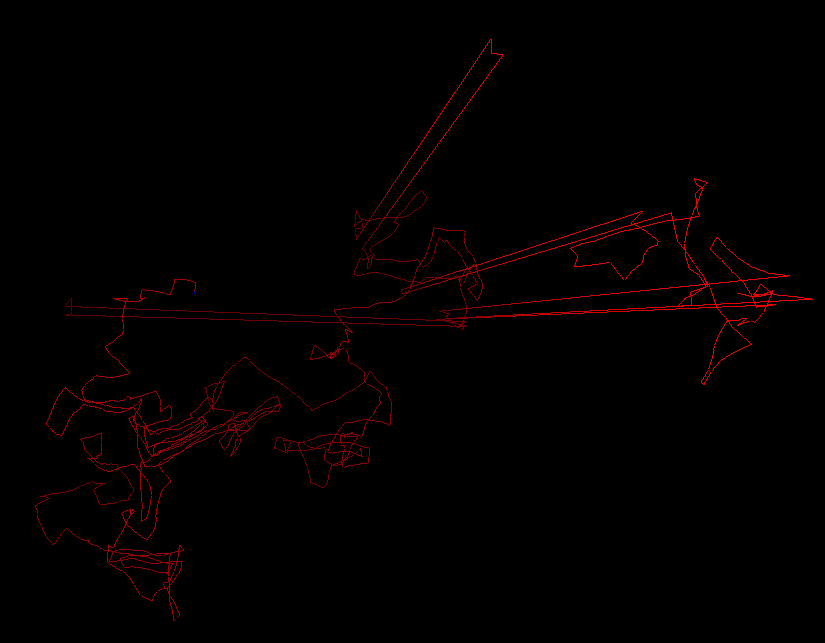}
	\end{subfigure}
    \caption{Characteristic conformations of growing chains in the (a) slow reaction regime, and (b) fast reaction regime. Characteristic accessible spheres for the chains are marked by the dashed circles. (c) and (d) - corresponding conformations from computer simulations of a single chain growth (see Computer Simulation Results section and Supporting Information).}
    \label{regimes12}
\end{figure}

If the fraction of initiators is relatively small, the chains can become long enough to collapse in the system, which is below the coil-globule transition temperature for the long chains. We suppose, that the reaction rate controls the conformation of the long growing chain, and we confirmed this assumption by computer simulations of a single chain growth (Fig. \ref{singlechaingrowth}). The following two regimes are possible in the first phase. A growing chain in a poor solvent can resemble a globular Brownian particle if the reaction rate is small enough, so every newly attached monomer collapses to the already grown globule. The similar conformations are observed for the single chain growing with $r=10^{-3}$ (Fig. \ref{regimes12}c, \ref{singlechaingrowth}). This model is described in the subsection II.B.1. On the other side, if the growing process is relatively fast, the chain will not resemble a spherical globule, but a locally collapsed Gaussian chain. These conformations are observed for the single chain growing with $r=5\times10^{-1}$ (Fig. \ref{regimes12}d, Fig. \ref{singlechaingrowth}). We developed a simple model for describing such conformations in the subsection II.B.2.

\subsubsection{Slow reaction regime}
In this regime the growing chain is in the globular state, so in the simplest approximation we consider the chain as a spherical Brownian particle with a time-dependent radius $R(t)\propto N^{1/3}(t) = (rt)^{1/3}.$ If we take into account the diffusion motion of this particle, the virtual sphere of radius $R^*(t)$ is accessible to it (Figure \ref{regimes12}a). We estimate $(R^*(t))^2$ as the mean square displacement of the center of mass of the particle over time $t$, which obeys the following well-known equation $\frac{d(R^{*}(t))^2}{dt} = 6D$, where $D$ is the diffusion coefficient $D = \frac{k_BT}{6\pi\eta_sR(t)}$ (here $\eta_s$ is the viscosity of the solvent). After integration of the equation we obtain the following scaling relation $R^*(t)\propto r^{-1/6}t^{1/3}=r^{-1/2}N^{1/3}$. Brownian particles should diffuse for at least the average distance between them before collision ($2(\frac{3}{4\pi}\frac{1}{c\rho})^{1/3}$), where $\rho$ is the particle density in the system) to have a chance to collide and coagulate, starting to form entanglements. Therefore, $R^*(N_c)$ at least should be equal to $R^*(N_c)=2(\frac{3}{4\pi}\frac{1}{c\rho})^{1/3}$, and the lower bound for $N_c$ in this regime of the slow reaction is given by Equation \ref{eq:2}.

\begin{equation}
\label{eq:2}
    N_c^{(i)}\propto r^{3/2}c^{-1}
\end{equation}

This formula with exact coefficients is given in the Application 2 (Fig. \ref{ncranalytical}).

\subsubsection{Fast reaction regime}
In this regime we assume that the chain grows fast enough and resembles a Gaussian walk of collapsed blobs, as shown on the Figure \ref{regimes12}b. Hydrodynamic radius of such a state is much larger, than the radius of a spherical globule of the equivalent chain length, so we do not take into account the Brownian motion of the whole chain in this regime. We suppose that the size distribution of blobs along the chain can be very non-trivial, but this distribution and the average size of the blob is determined by the interplay between the processes of growth and collapse of the chain in a poor solvent. 

Following this idea, we developed a simple approach to characterize the minimal size of the blob in the system. We assume, that if the polymerization reaction is relatively fast, the new monomers do not continuously collapse on the single previously synthesized globule, but firstly form a long non-collapsed subchain. This subchain starts to collapse afterwards according to the pearl-necklace mechanism, and, therefore, the blobs form along the chain. Following this model of the fast chain growth, we derive the minimal blob size $b$ accurately in the Application 3. We only state here, that the time of growth of the subchain, which forms the blob with the minimal size $b$, is of the order of collapse time of this subchain. The blob growth time is $t_b = N_b/r$, where $N_b$ is the average number of monomers in a blob. The blob collapse time can be estimated as $\tau_b \propto N_b^2$ \cite{kuznetsov1996kinetic,kuznetsov1996equilibrium,rostiashvili2003collapse,kikuchi2005kinetics}. Assuming that $b \propto N_b^{1/3}$, these relations give us the scaling relation, how the minimal blob size depends on the reaction rate in this model $b \propto r^{-1/3}$. Detailed derivation of this expression with exact coefficients is given in the Application 3.

For simplicity, we estimate the average chain size neglecting the fact that there are blobs larger than the blob with the minimal size (further: blob), representing the chain as a Gaussian walk of blobs size $b$ (Fig. \ref{regimes12}b). In this crude approximation, the average number of blobs in the chain length $N$ is $n_b = N/N_b \propto r^2t$. The end-to-end distance of the Gaussian chain consisting of $n_b$ blobs of radius $b$ is $R(t) = bn_b^{1/2} \propto r^{2/3}t^{1/2}=r^{1/6}N^{1/2}$. We can determine $N_c$ as we did in the subsection II.A, $cR^3(N_c)\propto1$. Therefore, the average chain length before the interchain contact in the fast reaction regime is given by the Equation \ref{eq:3}.

\begin{equation}
\label{eq:3}
    N_c^{(ii)}\propto r^{-1/3}c^{-2/3}
\end{equation}

\begin{figure}[h!]
    \centering
	\includegraphics[width=\linewidth,keepaspectratio]{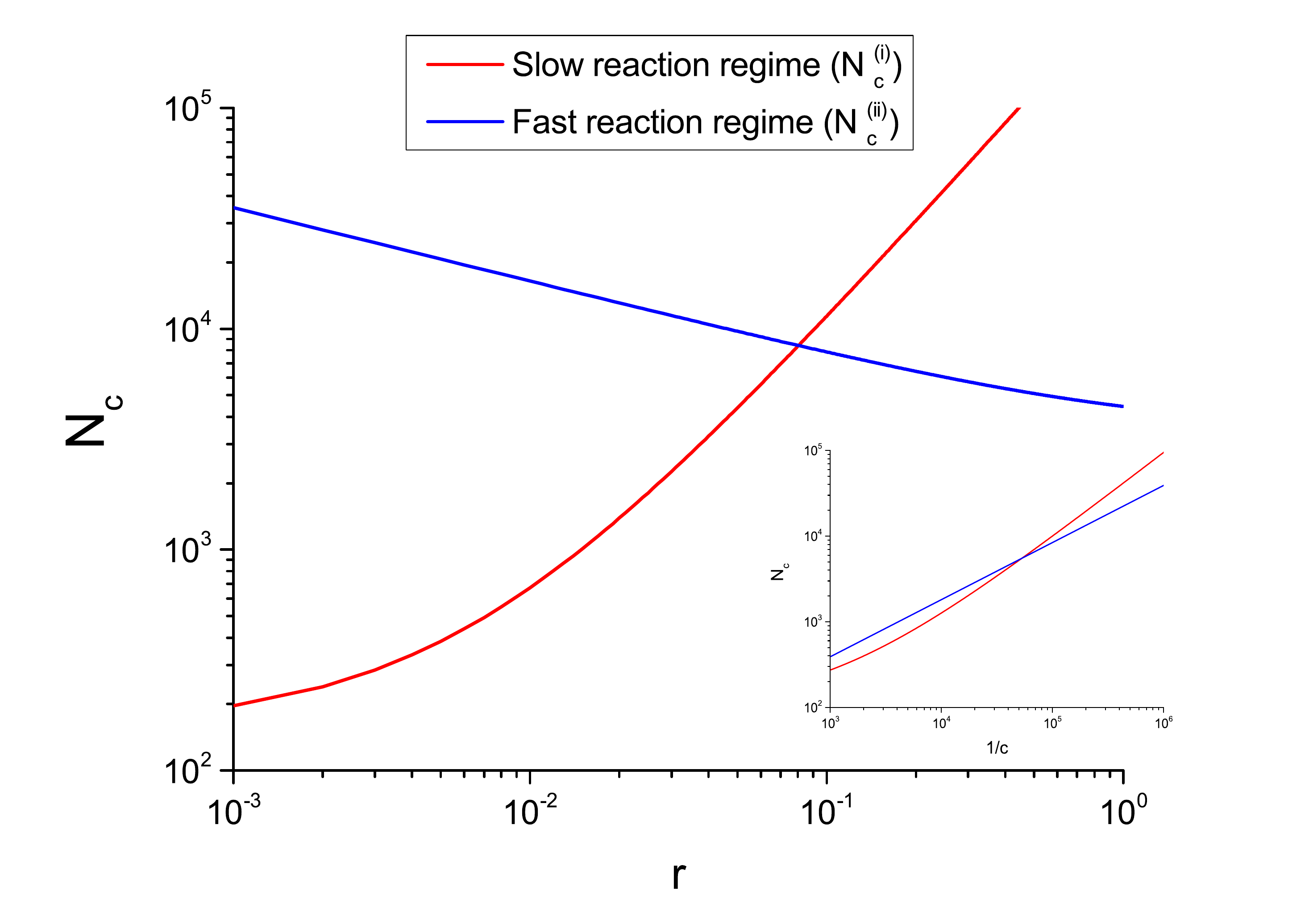}
	\caption{Dependencies of $N_c$ (maximum chain length reached in the first phase of the chain growth) on the reaction rate $r$ (at $c = 10^5$) and fraction of initiators $c$ (at $r = 0.08$, inset) in the slow and fast reaction regimes.}
    \label{ncranalytical}
\end{figure}

This formula with exact coefficients is given in the Application 3 (Fig. \ref{ncranalytical}).

\subsection{Formation of entanglements}
As we have described above, growing chains start to interact with each other and form entanglements after reaching the average length, which is a function of reaction rate and fraction of initiators: $N_c(r,c)$. In this section, we develop a simple model to describe evolution of the average entanglement length $N_e(t)$. More precisely, we will describe the function $N_e(N)$, and since the relation $N=rt$ holds almost for the whole range of degrees of polymerization (Fig. \ref{conversion}), we imply that the function $N_e(N)$ describes the time evolution of the entanglement length in the system as well.

First, we make assumptions about the mechanism of entanglements formation during growth of the chains. Let us denote the average number of entanglements per chain as $Q = \frac{N}{N_e(N)}$ and assume the following:

(1) Entanglements are formed spontaneously only when the growing end of one chain grows through the space, where another chain is present. We neglect effects of entanglement/disentanglement caused by reptation of chains or diffusion in the solution.

(2) If the chain length $N$ increases its length on $N_e(N)$ new beads (i.e. on the current average entanglement length in the system), a constant average number of new entanglements $Q^*$ per chain will be formed. $Q^*$ may depend on $r$ and $c$, but remains constant during the whole polymerization process with fixed $r$ and $c$.

Following our initial assumption, entanglements start to form when the average chain length is $N_c$. The average number of entanglements at this moment becomes $Q(N_c) = 1$, and, therefore, the average entanglement length in the system is $Ne(N_c) = N_c/2$. Following the assumption (2) described in the previous paragraph, when the chain increases its length on $N_e(N_c)$ beads, the average number of entanglements increases on $Q^*$, and the average entanglement length decreases correspondingly:

\begin{equation}
\label{eq:4}
\begin{cases}
N_c \to N_1 = N_c+N_e(N_c), \\
Q(N_c) \to Q(N_1) = Q(N_c) + Q^*, \\
N_e(N_c) \to N_e(N_1) = N_1/(1+Q^*).
\end{cases}
\end{equation}

If we denote the average chain length before the next recalculation of $N_e$ according to the rule (2) as $N_j$, we can write the Equation \ref{eq:4} in more general form:

\begin{equation}
\label{eq:5}
\begin{cases}
N_0=N_c;\\
Q(N_0)=1;\\
N_e(N_0)=N_c/2;\\
N_j \to N_{j+1} = N_j+N_e(N_j), \\
Q(N_j) \to Q(N_{j+1}) = Q(N_j) + Q^*, \\
N_e(N_j) \to N_e(N_{j+1}) = N_{j+1}/Q(N_{j+1})=\frac{N_j+N_e(N_j)}{Q(N_j) + Q^*}.
\end{cases}
\end{equation}

This recalculation process continues until the chains reach the maximum possible average length $N_j\approx1/c$. As we discuss in the Computer Simulation Results section, the iterative recalculation procedure (Eq. \ref{eq:5}) gives the power law decreasing function (Equation \ref{eq:6}).

\begin{equation}
\label{eq:6}
\begin{cases}
N_e(N) \propto N_c^{k+1} N^{-k}\\
N>N_c
\end{cases}
\end{equation}

The exponent $k\geq0$ depends only on $Q^*$ and varies from $k=0$ (for $Q^* = 1$, leading to $N_e(N) = N_c/2$) to $k=1$ (for $Q^* = 2$) and larger values of $k$ (for larger $Q^*$). 
According to the computer simulation data, the $Q^*$ value varies from $Q^* = 1.2$ for $r=0.001$ to $Q^* = 1.7$ for $r=0.5$ (Fig. \ref{nen_sim}b). One of the most important features of the Equation \ref{eq:6} is that the $N_c(r,c)$ controls the entanglement length decrease, and, in particular, the entanglement length after polymerization (i.e. $N_e(N=1/c)\propto N_c^{k+1} c^{k}$. 

\subsection{Model predictions}
We have assumed in our model, that there are two different regimes of the chain growth for the small fraction of initiators $c$. The first regime is the growing globular Brownian particle, which is formed by the consecutive collapse of each new monomer to the globule (possible for very small $r$). The second regime is the growth of locally collapsed Gaussian chain, in which the minimal blob size can be estimated assuming that for very large polymerization rates $r$ there is no consecutive collapse of each new monomer to the already synthesized chain, but instead new monomers first form a relatively long non-collapsed subchain, which then collapses according to the pearl-necklace mechanism. The behavior of the growing chain may be much more complex for the intermediate reaction rates. However, we naturally assume, that $N_c(r,c)$ is a continuous function of both arguments. In this work, we implement a simple interpolation method for constructing the unified dependency $N_c(r,c)$ for all possible reaction rates. We continue both dependencies $N_c^{(i)}(r,c)$ and $N_c^{(ii)}(r,c)$ until their intersection on the line defined by the equality $N_c^{(i)}(r,c)=N_c^{(ii)}(r,c)$.

As we have emphasized in the subsection II.A, the chain collapse is possible only for the long enough chains in our system. We estimated the minimal collapsing chain length $N_{min}\approx100$ beads. If $N_c\approx N_{min}$, we suppose, that quite complex effects connected with the beginning of the chain collapse are possible. In our simple approach, we do not estimate how these effects affect the true $N_c$ value. We simply state, that if $N_c$ value derived from the Equation \ref{eq:1} is less than $N_{min}$, we predict $N_c$ from the Equation \ref{eq:1}. In the opposite case, we predict the $N_c$ value from either slow or fast reaction regime.

\begin{figure}[h!]
    \centering
	\includegraphics[width=\linewidth,keepaspectratio]{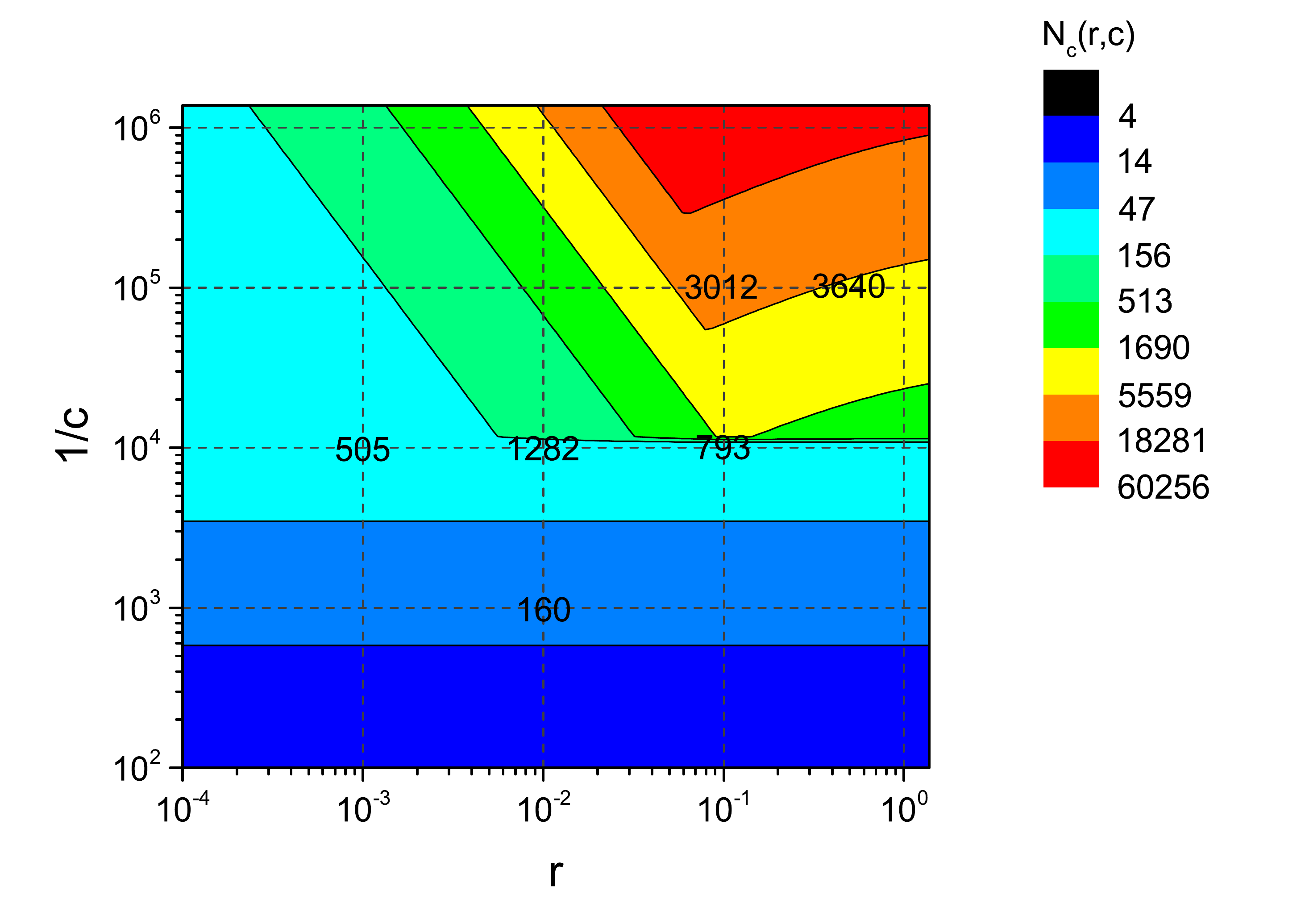}
	\caption{Analytically computed $N_c(r,c)$ dependency. Numbers on the plot represent the $N_c$ values (peaks of the $N_e(N)$ dependencies) obtained from MD simulations.}
    \label{ncne}
\end{figure}

Based on this simple approach, we can build the unified $N_c(r,c)$ dependency (i.e. the maximal possible average chain length before it starts to form entanglements, Fig. \ref{ncne}). Numerical coefficients, which are needed for exact calculation of the Fig. \ref{ncne} depend on the parameters of the system, and are taken from the computer simulation model presented in the next section.

\section{Computer simulations}
Experimental studies consider partial crystallization of growing chains as a main factor to synthesize low-entangled melts. Therefore, in our simulation studies we include effect of chain crystallization as a natural feature of polymer chain under consideration.

\subsection{Molecular dynamics model}
We performed computer simulations of polymerization process and studied the resulting polymer melt using molecular dynamics (MD). We used a simplified model, where catalysts (reaction initiators) were dissolved in liquid of monomers, and this liquid was a poor solvent to the growing chains. This "liquid of monomers" imitated the poor solvent mixed with injected monomers in the reaction well. Initiators, single monomers and polymerized monomers were represented as particles of different types (I, M and P, respectively). The length unit was chosen $\sigma=0.52nm$, and the dimesionless time unit is denoted as $\tau$. All particles were interacting by LJ96 potential with the following parameters: $\epsilon=1.511k_BT$ and $\sigma_0=0.89\sigma$, which were used for simulations of polymer melt crystallization in the work \cite{luo2009coding}. The potential for all interactions of non-bonded beads, except for the solvent-solvent interactions, was cut off at the distance $R_{cut}=1.02\sigma$, where the aforementioned LJ96 potential reaches its minimum \cite{luo2009coding}. The LJ96 potential for solvent-solvent interactions was cut off at $R_{cutSS}=0.9\sigma$ to model the poor solvent conditions for long chains (see Supporting Information). Bonds were simulated by the following harmonic potential: $1352.0\times(r-0.5\sigma)^2$, $r$ is the distance between centers of beads. The angle potential was tabulated to model polyethylene, this potential was used in the works Ref. \cite{luo2009coding,meyer2002formation}, and was applied to model crystallization of the chain into lamellar structures. All simulations were carried out in NPT ensemble at particle density $\rho=0.9$, temperature $T=0.5$ and pressure $P=8.0$, damping rates for pressure and temperature were equal to $10$ MD time steps. The aforementioned conditions ensured collapse of long chains along with their simultaneous crystallization (see Supporting Information and Ref. \cite{luo2009coding}).

We used "mesoscale chemistry" approach to model the polymerization reaction in our simulations. Initiators and monomer particles were initially marked with valence of 1 and 2 ($I.^*$ and $:M$), respectively, and each initiator particle was marked as an active center (asterisk in all following notations in this section). Two reactions were allowed:

(1) Initialization reaction, during which the initiator with active mark and a monomer with free valence and no active mark form a bond. The active mark passes to the monomer, and valences of both particles decrease by one. The type of monomer particle changes to ‘polymerized”: $I.^* + :M \to I-P.^*$.

(2) Polymerization reaction, during which a polymerized monomer with a free valence and active mark and the second monomer with a free valence and no active mark form a bond. The active mark passes to the second monomer, and valences of both particles decrease by one: $-P.^* + :M \to -P-P.^*$.

Every particle marked as active center was checked for having neighbors fitting the reaction rules inside radius $R_{cut} = 1.02\sigma$ every $200$ MD steps, which is equal to 1 time unit. If there were more than one neighbor available for reaction, a random one was chosen. Then bond was formed between these particles, with the probability $p$. The product of time interval and reaction probability determined the effective reaction rate $r=p$ (in reverse time units) and varied from $r=10^{-3}$ to $r=5\times10^{-1}$.

All simulations were performed in LAMMPS package \cite{plimpton1995fast} with modified fix\_bond\_create module which implements multiple reactions. Modified LAMMPS module, run script sample and resulting synthesized 39000-particle microgel structure (with sol fraction cut out) are available for download (LAMMPS module for multiple reactions with samples, Laboratory of Microstructured Polymer Systems. http://polly.phys.msu.ru/~rudyak/lammps\_fix\_bond\_create.html).

\subsection{Computer simulation results}
We have performed computer simulations to validate our theoretical approach. First we have estimated numerical constants in our system, which are needed for exact calculation of the $N_c(c,r)$ (Fig. \ref{ncranalytical} and Fig. \ref{ncne}). From mean-squared displacement (MSD) measurements we have obtained the value of $D(a) = \frac{k_BT}{6\pi\eta_sa}$ - diffusion coefficient of a bead with radius $a=0.5\sigma$ (equilibrium bond length in our simulations). MSD measurements were carried out in the liquid of beads under conditions described in the previous subsection, LJ96 potential was cut off at $R_{cut}=1.02\approx2a$ - effective radius of the bead. In this system $MSD(t = 1 time unit = 200 MD steps) = 6D(2a) = \frac{k_BT}{\pi\eta_s2a} = 0.2138 \sigma^2$. Therefore, $\frac{k_BT}{\pi\eta_sa} = 0.4276 \frac{\sigma^2}{timeunit}$. The second coefficient, which is needed to be estimated, is $F$ - coefficient of proportionality in the law for the chain collapse time $\tau_{coll.subch.} = FN^2$ (see Application 3 for details). For chain length of the order of magnitude $N\propto10^2$ beads collapse time is of the order of magnitude $\tau_{coll.subch.}\propto 10^4$ time units (Fig. \ref{coilglob}). Therefore, $F\propto10^0$ time units. To estimate $F$ more precisely, we needed to carry out simulations of collapse of chains, which differ by several orders of magnitude in length. However, due to crude approximations used for building $N_c(r,c)$ dependency, and weak dependency of $N_c^{(ii)}$ on $F$ ($N_c^{(ii)}\propto F^{-1/3}$), precise estimation of the $F$ value is not so important, so we used the value $F=1$ time unit (Fig. \ref{ncranalytical}, Fig. \ref{ncne}). Finally, we used the value $n_p=8$ beads to construct the $N_c(r,c)$ dependency (Fig. \ref{ncne}). We estimated this value from $R(s)$ dependencies in the single chain and many chain systems in the beginning of chain growth (Fig. \ref{singlechaingrowth} and \ref{rs}).

We used the following values for the fraction of initiators in our simulations: $c=10^{-3}$ ($184$ initiators, $184\times 10^3$ particles total), $c=10^{-4}$ ($18$ initiators, $184\times 10^3$ particles total) and $c=10^{-5}$ ($9$ initiators, $9\times 10^5$ particles total). These values were chosen to describe the system in the large $c$ limit, in the transition area from the large and small $c$ limit, and in the small $c$ limit, respectively (Fig. \ref{ncne}).

\begin{figure}[h!]
    \centering
	\begin{subfigure}{0.2\textwidth}
	\includegraphics[width=\linewidth,height=\textheight,keepaspectratio]{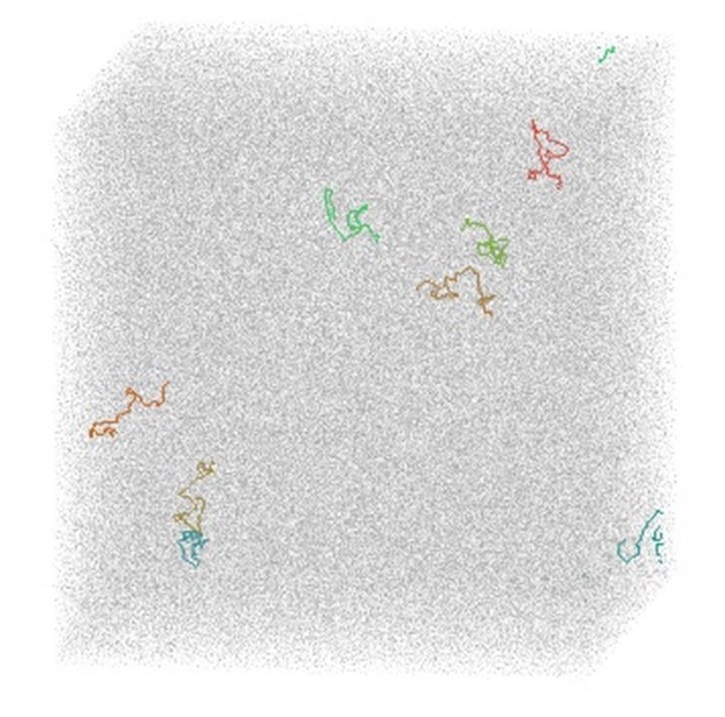}
	\end{subfigure}
	\begin{subfigure}{0.2\textwidth}
	\includegraphics[width=\linewidth,height=\textheight,keepaspectratio]{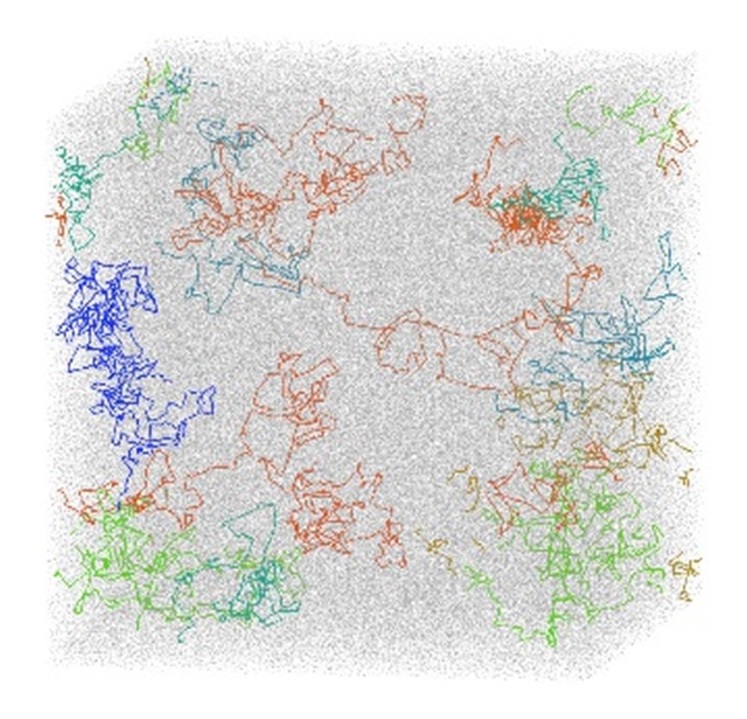}
	\end{subfigure}
	\begin{subfigure}{0.2\textwidth}
	\includegraphics[width=\linewidth,height=\textheight,keepaspectratio]{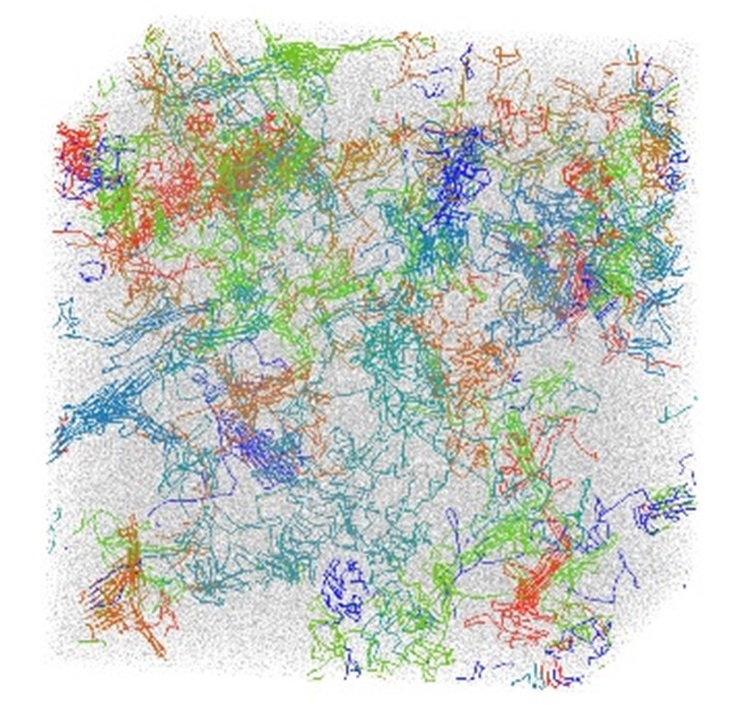}
	\end{subfigure}
	\begin{subfigure}{0.2\textwidth}
	\includegraphics[width=\linewidth,height=\textheight,keepaspectratio]{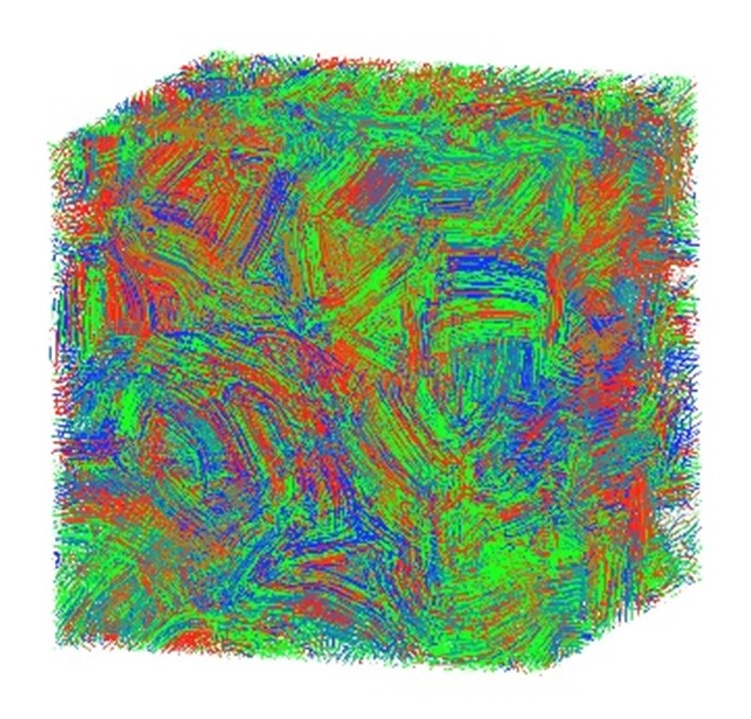}
	\end{subfigure}
    \caption{Snapshots taken during chain growth in the system with $c=10^{-5}$ and $r=5\times10^{-1}$.}
    \label{snapshots}
\end{figure}

We built the sequence of snapshots during chain growth in the system with $c=10^{-5}$ and $r=5\times10^{-1}$ as an illustration of the chain growth process in our systems (Fig. \ref{snapshots}). We can see, that there is indeed the phase of individual growth of the chains, where chains exhibit Gaussian conformations (Fig. \ref{snapshots}a, \ref{snapshots}b) (as predicted by the fast reaction regime model for these values of $c$ and $r$ Fig. \ref{ncne}). This phase is followed by entanglement of chains and melt formation (Fig. \ref{snapshots}c, \ref{snapshots}d). See Supporting Information for analysis of conformations during chain growth (Fig. \ref{rs}) and snapshots of the system of slowly growing chains (Fig. \ref{snapsslow}).

\begin{figure}[h!]
    \centering
	\begin{subfigure}{0.45\textwidth}
	\includegraphics[width=\linewidth,height=\textheight,keepaspectratio]{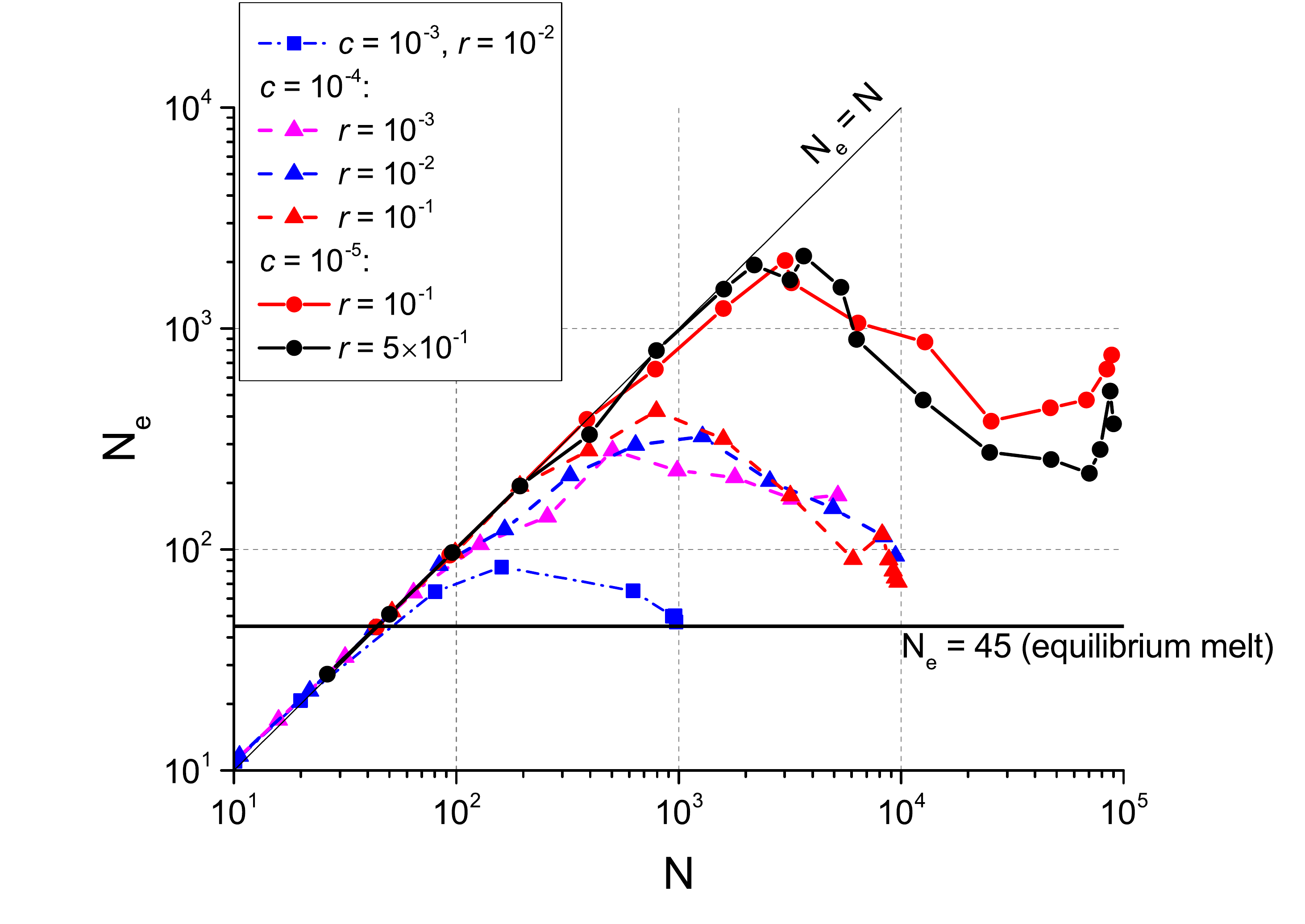}
	\end{subfigure}
	\begin{subfigure}{0.45\textwidth}
	\includegraphics[width=\linewidth,height=\textheight,keepaspectratio]{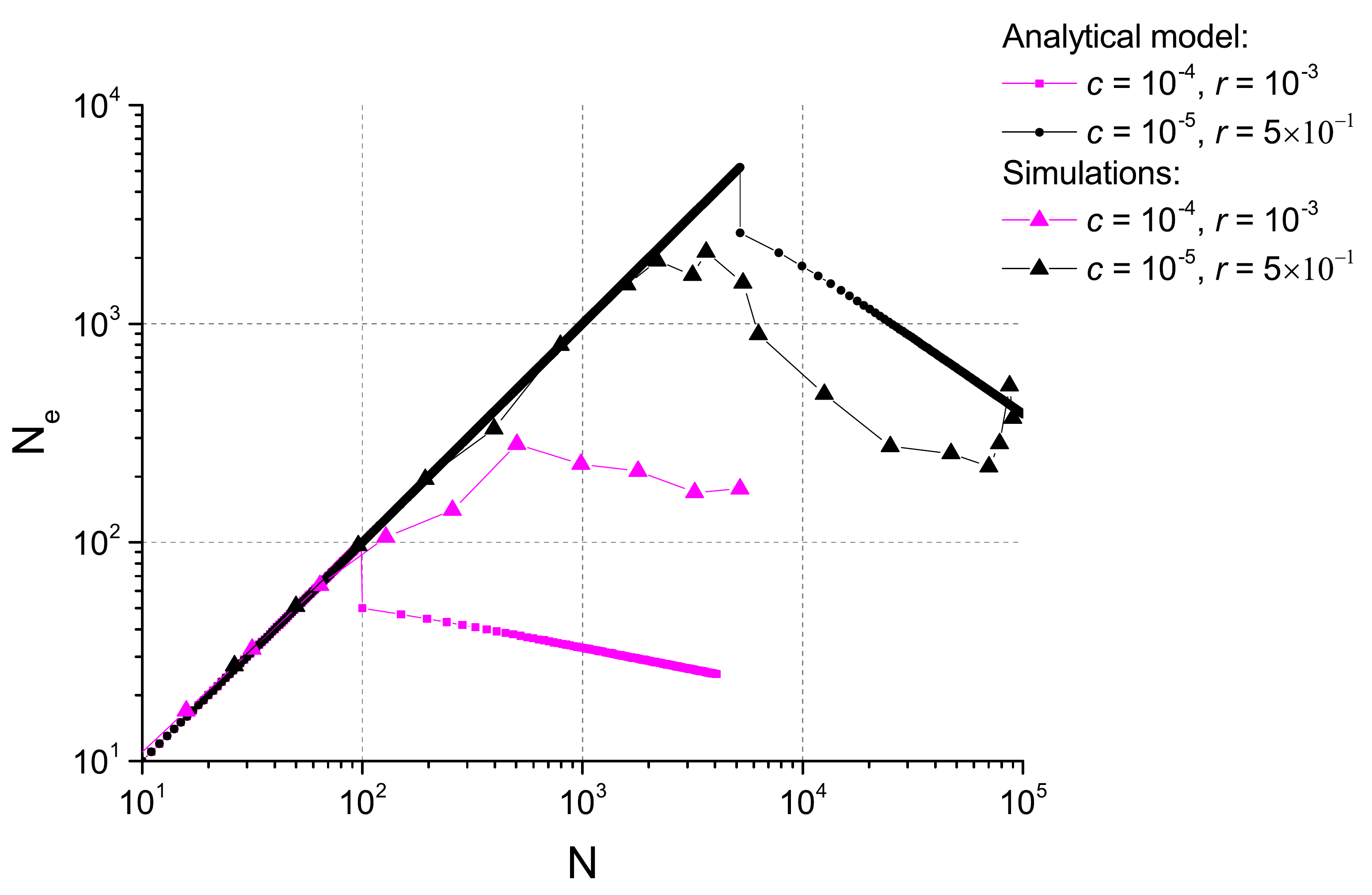}
	\end{subfigure}
    \caption{The average entanglement length $N_e$ as a function of the average chain length $N$ for various systems. a) Coarse grained MD data. Thin black line represents the $N_e=N$ dependency (expected in the first phase of chain growth). Thick horizontal black line represents the average entanglement length in the melt crystallized from the equilibrated melt (see Supporting Information). b) Comparison of $N_e(N)$ dependencies derived from our analytical theory and MD data in the two cases of slow and fast reaction.}
    \label{nen_sim}
\end{figure}

To validate the model more quantitatively, we built $N_e(N)$ dependencies for the 6 systems with different reaction rates or fractions of initiators (Fig. \ref{nen_sim}a). First, we observed qualitative agreement with the theoretically predicted picture: there is a phase with $N_e=N$ dependency, i.e. chains grow independently and do not form entanglements. This phase ends at some well-defined value $N=N_c$, and $N_e$ starts to decrease with $N$. We see, that $N_c$ strongly depends on $c$ and systems with smaller $c$ and the same $r$ have larger $N_e$ in the end of polymerization, i.e. when $N\approx1/c$. $N_e(N=1/c)$ for small $c$ is much larger than the entanglement length in the melt crystallized from the equilibrated melt (see Supporting Information). The latter finding also agrees with experimental studies \cite{pandey2011heterogeneity}. Interestingly, that we observed increase of $N_e$ at large $N$ for the systems with $c=10^{-5}$ (Fig. \ref{nen_sim}a). This behavior was not predicted by our analytical theory and is observed probably due to growing of the chain ends in the monomer-rich areas inside a dense crystallized melt, where the chain end can not form entanglements with other chains due to their crystallized state. However, this effect does not lead to qualitatively different $N_e(N=1/c)$ values from the analytical predictions (Fig. \ref{nen_sim}b).

We built also the theoretically predicted $N_e(N)$ dependencies for two systems: $r=10^{-3}$, $c=10^{-4}$ (further: first system) and $r=5\times10^{-1}$, $c=10^{-5}$ (further: second system) (Fig. \ref{nen_sim}b). $Q^*$ values, necessary for calculation of the second phase in the $N_e(N)$ dependency, were taken as follows: $Q^*=1.2$ and $Q^*=1.7$ in the first and the second system, respectively. Therefore, the exponent $k$ in the dependency $Ne\propto N^{-k}$ is close to the value observed in the simulations: $k\approx 0.21$ in the first system and $k\approx 0.66$ in the second system. Figure \ref{nen_sim}b shows qualitative agreement of the theoretically predicted and simulated $N_e(N)$ dependencies (difference between predictions is less than one order of magnitude). We also investigated, why $k$ values are different in the first and the second systems. This can be due to the different states, where formation of entanglements occurs. Formation of entanglements in the first system occurs in the crystallized state, where all chains are the part of the single crystallized structure (Fig. \ref{snapsslow}). On the other side, formation of entanglements occurs in the solution of partly crystallized chains in the second system (with 500-fold larger reaction rate) (Fig. \ref{snapshots}). We have also observed differences in the chain folding in the melt after polymerization obtained with different reaction rates (Fig. \ref{rs}). Therefore, as we have suggested in our theoretical approach, value of $Q^*$ (or equivalently, the exponent $k$) characterizes the process of entanglements formation and is dependent on the reaction rate, but is constant during polymerization with fixed $r$ and $c$.

Finally, we performed scaling of the system with $c=10^{-4}$ and $r=10^{-1}$ to prove, that $N_e(N=1/c)$ value does not depend on the size of the system. We carried out simulations of the system with $9$ initiators and $9.2\times10^{4}$ particles in total. We proved that the chain conformations are similar and the $N_e(N\approx1/c)$ values coincide within the error for the small and large system, and both values are larger than $N_e\approx45$ in the equilibrium melt (Fig. \ref{sizeeffect}, \ref{nedistrcomp}).

\section{Conclusions and Discussion}
In our work, we investigated the process of melt formation by homogeneously catalyzed polymerization in a poor solvent with simultaneous crystallization of growing chains. We developed a simple analytical approach, which describes the process of entanglements formation. We compared the theoretical predictions with computer simulations of coarse-grained model of polyethylene.

(1) MD simulations supported our basic theoretical picture of the chain growth process, i.e. the existence of two phases of chain growth: independent chain growth (first phase), and entanglement of chains (second phase).

(2) MD data showed the existence of different chain morphology in the first phase predicted by the model: non-collapsed Gaussian conformation of short chains, globular and locally collapsed conformations of long chains.

(3) The difference of theoretically predicted values of $N_c$ (which determine the duration of the first phase $t_c=N_c/r$) and MD data is less than one order of magnitude for all studied systems.

(4) There are non-trivial predictions of the analytical model of the second phase of chain growth, and these predictions were confirmed by the simulations. First of all, the power law decrease of the average entanglement length with $N$ was observed in MD simulations for all systems. Secondly, decrease of the fraction of initiators $c$ (with fixed $r$) led to increase of $N_c$ and $N_e(N=1/c)$ in our MD simulations and experimental studies \cite{pandey2011heterogeneity}. This is a non-trivial behavior, also predicted by our theoretical approach.

However, our simple theoretical approach can not predict several phenomena observed in the simulations. First of all, the exponent $k$, which characterizes the power law decrease of $N_e$ with $N$ in our theory, is a phenomenological constant, which depends on the reaction rate, and, probably, on the fraction of initiators. Therefore, we could not build the analytical dependency of the average entanglement length in the end of polymerization $N_e|_{N=1/c}(r,c)$, as we did for $N_c(r,c)$ (Fig. \ref{ncne}). Secondly, our theoretical approach fails to predict $N_c$ quantitatively due to great simplification of the model: we do not take into account the non-spherical shape of the crystallized polymer globule (Fig. \ref{regimes12}c) and do not take into account the real detailed structure of the locally collapsed state (Fig. \ref{regimes12}d). Moreover, it is clear, that the $N_c(r,c)$ dependency constructed by the simple interpolation method (Fig. \ref{ncne}) can not quantitatively predict $N_c$ values. This method also leads to the non-physical dependency of the blob size on $r$ and $c$: in fact, there is a jump of the blob size on the line defined by the equality $N_c^{(i)}(r,c)=N_c^{(ii)}(r,c)$, and we consider this as the artefact of the method. Further research is needed to construct the $N_c(r,c)$ dependency taking into account the true physical behavior of the partly crystallized chain growing in a poor solvent.

Our theoretical approach predicts a non-trivial dependency of the $N_c(r,c)$ on $r$ in the small $c$ region (Fig. \ref{ncne}). We predict the existence of the "optimal" reaction rate, which leads to the system with the maximal possible $N_c$ for the given $c$. 
We were not able to test these predictions for the slow reaction regime due to computational restrictions (calculation of the full polymerization in the system with $c=10^{-5}$ and $r=0.5$ took 2 months of computational time).


\section{Acknowledgments}
The research is carried out using the equipment of the shared research facilities of HPC computing resources at Lomonosov Moscow State University \cite{voevodin2019supercomputer}. The research of Artem Petrov is supported partly by the grant from the Foundation for the advancement of theoretical physics "Basis''. The reported study was funded by RFBR according to the research project \# 18-03-01087.

\section{Application 1}
If the chain is short enough, it does not collapse in our system, but instead forms a Gaussian conformation with persistence length $n_p\approx10$ beads. In the system of many dissolved initiators, we can draw a virtual sphere around each initiator such that the spheres become closely packed. The radius of the spheres is $R_{sph} = (\frac{3}{4\pi}\frac{1}{c\rho})^{1/3}$, where $\rho$ is the particle density in the system. Gaussian chains may start to intersect and, therefore, form entanglements, when end-to-end distance becomes equal to $R_{sph}$. Hence, $a(N_cn_p)^{1/2}=(\frac{3}{4\pi}\frac{1}{c\rho})^{1/3}$ and we obtain the Equation \ref{eq:nclarge}.

\begin{equation}
\label{eq:nclarge}
\begin{cases}
    N_c(c) = \frac{1}{a^2n_p}(\frac{3}{4\pi}\frac{1}{c\rho})^{2/3}\propto c^{-2/3}\\
    N_c(c)<N_{min}\approx 100
    \end{cases}    
\end{equation}

\section{Application 2}
As we have previously discussed, if the polymerization rate is small enough, the growing chain in the poor solvent resembles a collapsed globule. Neglecting the real shape of the growing crystallized globule, we describe the growing chain as a spherical Brownian particle with time-dependent radius. A simple assumption is that $N_{c}$ (the average number of monomers in a globule, when globules collide at least pairwise) should be at least the average number of monomers in a growing globule, when its average displacement is $2R_{sph}$ (i.e. the average distance between particles before their coagulation) (further: Assumption 1).

It is a well known expression for the mean-squared displacement of a spherical Brownian particle $\frac{d(R^{*}(t))^2}{dt} = 6D$, where $D$ is the diffusion coefficient $D = \frac{k_BT}{6\pi\eta_sR(t)}$ (here $\eta_s$ is the viscosity of the solvent). As we roughly represent the growing chains as spherical globules, we can simply write $R(t) = a(N+N_{min})^{1/3}$. Next we assume, that the degree of polymerization grows linearly with time, $N = rt$. Therefore, we have an equation for the chain displacement $d(R^{*}(t))^2 = \frac{k_BT}{\pi\eta_sa(r(t+t_{min}))^{1/3}}dt$, $t_{min}$ is the synthesis time of the chain length $N_{min}$. If we integrate it from t=0 to t,we obtain the expression $(R^{*}(t))^2 = \frac{3k_BT}{2\pi\eta_sar^{1/3}}[(t+t_{min})^{2/3}-t_{min}^{2/3}]$. Following the aforementioned Assumption 1, we obtain the expression for the lower bound of $N_{c}$ in the slow reaction regime (Equation \ref{eq:nc1}).

\begin{equation}
\label{eq:nc1}
\begin{split}
    N_c^{(i)}=[(2R_{sph})^2{\frac{2\pi\eta_sar}{3k_BT}}+N_{min}^{2/3}]^{3/2} = \\ [4(\frac{3}{4\pi}\frac{1}{c\rho})^{2/3}{\frac{2\pi\eta_sar}{3k_BT}}+N_{min}^{2/3}]^{3/2}\propto r^{3/2}c^{-1}
    \end{split}
\end{equation}

\section{Application 3}
If polymerization rate is large enough, the chain will resemble a locally collapsed Gaussian chain. We assume, that in this regime the newly attached monomers will no longer consequently collapse on the single globule, but will form long non-collapsed subchain, which will start to collapse afterwards. Let us estimate the minimal size of the crumpled region in this model, $b = aN_{b}^{1/3}$.

Various theoretical works, as well as simulation results suggest \cite{kuznetsov1996kinetic,rostiashvili2003collapse,kikuchi2005kinetics}, that the time sufficient for total collapse of the chain length $N$ is proportional to the squared length of the chain, $\tau_{coll.subch.}\propto N_{subch.}^2$, if the hydrodynamic interactions are not included. We will further employ this scaling, as we assume, that the length of the Gaussian subchain is not very large, so the hydrodynamic interactions do not affect the collapse kinetics. We denote the coefficient in the equation for collapse time as $F(T,\chi,microsc.param.)$, or simply $F$. This coefficient depends on solvent quality, temperature, and microscopic parameters of the chain potential \cite{kuznetsov1996kinetic}. We note, that for stiff polymers there is an additional term $\propto N_{subch.}^{1.3}$, which is added to the $\tau_{coll.subch.}$ (derived in the self-consistent Gaussian approach \cite{kuznetsov1996equilibrium}). We will not include this term in the further equations, as it is much smaller, that the term $\propto N_{subch.}^{2}$.

We assume, that the degree of polymerization grows linearly with time $N_{subch.}=rt_{synth.subch.}$. Hence, the collapse time of the newly synthesized subchain grows as $\tau_{coll.subch.}= F(rt_{synth.subch.})^2$. $\tau_{coll.subch.}$ grows as a quadratic function of $t_{synth.subch.}$. Therefore, if $t_{synth.subch.}$ is small enough, $\tau_{coll.subch.}$ is much less than $t_{synth.subch.}$. This fact breaks in the point determined by Equation \ref{eq:tau}.
\begin{equation}
\label{eq:tau}
    \tau_{coll.subch.}\approx t_{synth.subch.}
\end{equation}

Equation \ref{eq:tau} gives the value of $N_b$ - the characteristic maximum scale of length of a subchain, which collapses much faster than is synthesized: $N_b=1+(Fr)^{-1}$. Therefore, if collapse of a newly synthesized subchain has started (the subchain should be longer than $N_{min}$ to be able to collapse and much longer than $N_b$), the blob length $\propto N_b$ will form in a time $\tau<<\frac{N_b}{r}$. During this process additional subchain length $r\tau$ grows, but $r\tau<<N_b$. Hence, the newly synthesized subchain resembles a sequence of blobs size $b$, and a small new subchain length $r\tau<<N_b$, which is the starting point for the next subchain. Therefore, there is always a blob length $\propto N_b$ on the growing end of the chain, and it is the minimal crumpled subchain length in the grown chain (since blobs can merge together in the older parts of the chain). Assuming the spherical shape of the blobs, we obtain the following expression for $b$ (Equation \ref{eq:b}).
\begin{equation}
\label{eq:b}
    b=a(1+(Fr)^{-1})^{1/3}
\end{equation}

$b$ tends to the single monomer size $a$ in the limit of $r\to \infty$, which is expected for the chain being synthesized infinitely fast.

Assuming that all blobs in the chain have size $b$ (see subsection 2.2), we have the expression for the end-to-end distance of the Gaussian chain of blobs: $R^*(t)=b\sqrt{\frac{N}{N_b}}=a(1+F^{-1}r^{-1})^{-1/6}\sqrt{N}$. Since $N_c$ is defined as the number of monomers in the chain, which has end-to-end distance equal to the $R_{sph}$ value, we have the following equation for $N_c$ in the fast reaction regime (Equation \ref{eq:nc2}).

\begin{equation}
\label{eq:nc2}
    N_c^{(ii)}=c^{-2/3}(1+F^{-1}r^{-1})^{1/3}\frac{1}{a^2(\frac{4\pi}{3}\rho)^{2/3}}
\end{equation}

\section{Supporting Information}
\subsection{Equilibrium melt preparation and crystallization}
\begin{figure}[h!]
\centering
  \includegraphics[width=0.5\linewidth]{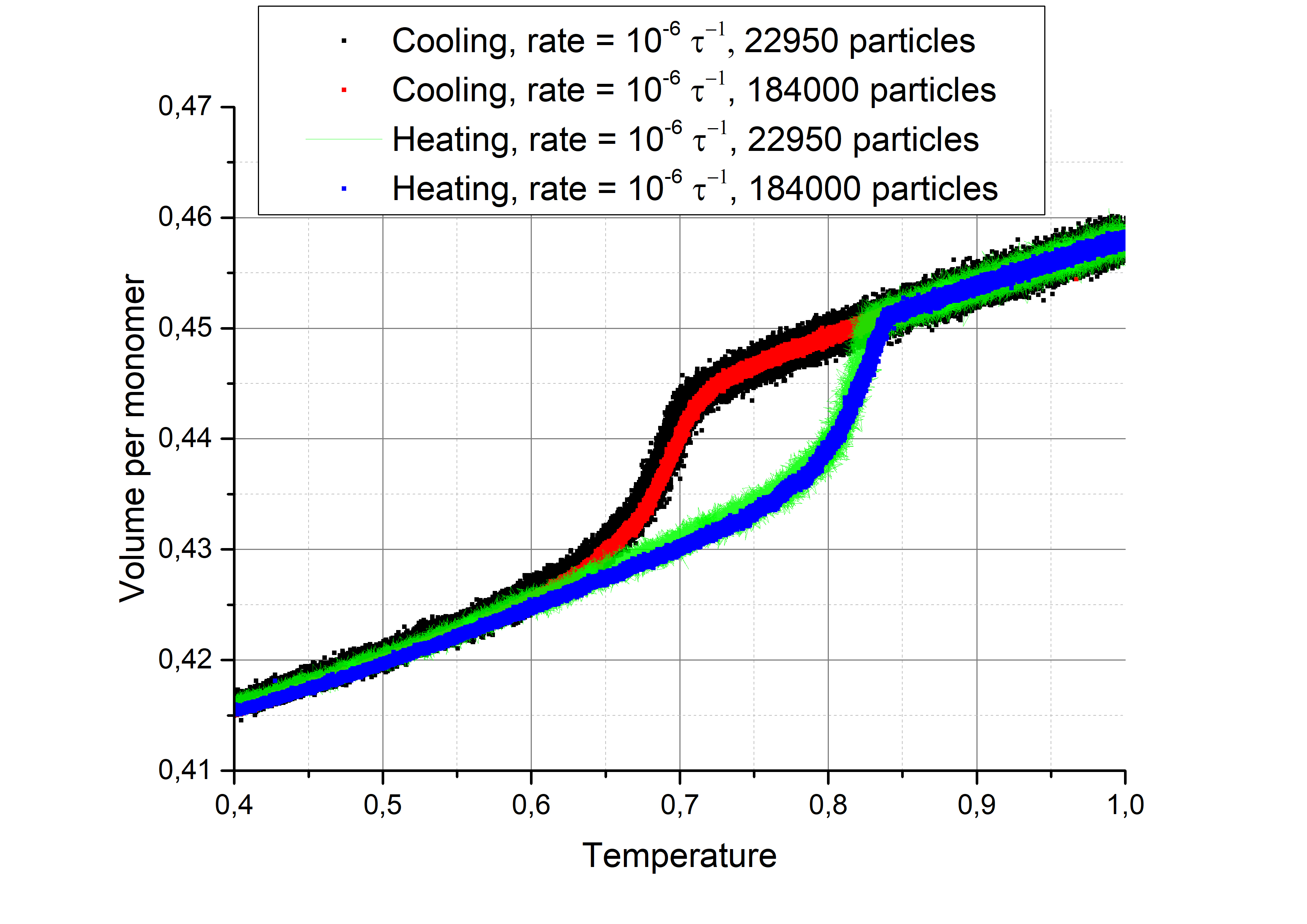}
\caption{Temperature dependencies of volume per monomer during cooling and heating with constant rate $10^{-6} \tau^{-1}$. Cooling was carried out from the equilibrium melt state. Equilibrium melt was obtained by synthesis described in the main article in the following two systems: the "smaller" system included $22.95\times 10^3$ particles and 23 initiators, and the "larger" system included $184\times 10^3$ particles and 184 initiators. Synthesis was carried out with reaction rate $r= 10^{-2}$ and temperature $T=1.0$.}
\label{crystallization}
\end{figure}

\begin{figure}[h!]
    \centering
	\includegraphics[width=0.5\linewidth]{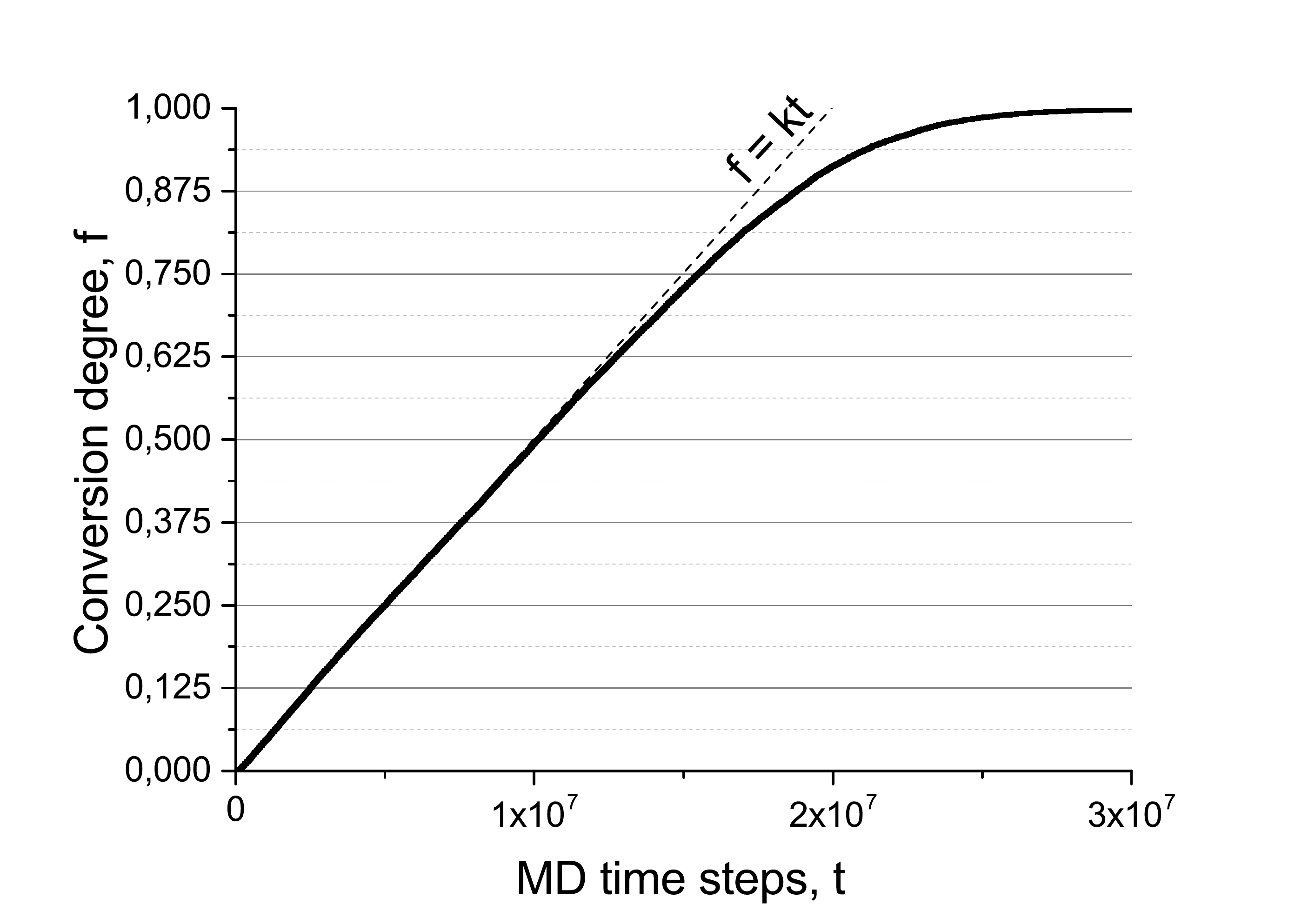}
	\caption{Conversion degree $f=cN$ (i.e. the fraction of initiators $c$ multiplied by the average chain length $N$) versus time in the "smaller" system during equilibrium melt synthesis. $k=cr$. We see, that the dependency $f=crt$ (i.e. $N=rt$) during synthesis holds up to $f\approx0.6$ perfectly.}
    \label{conversion}
\end{figure}

\begin{figure}[h!]
\centering
  \includegraphics[width=0.5\linewidth]{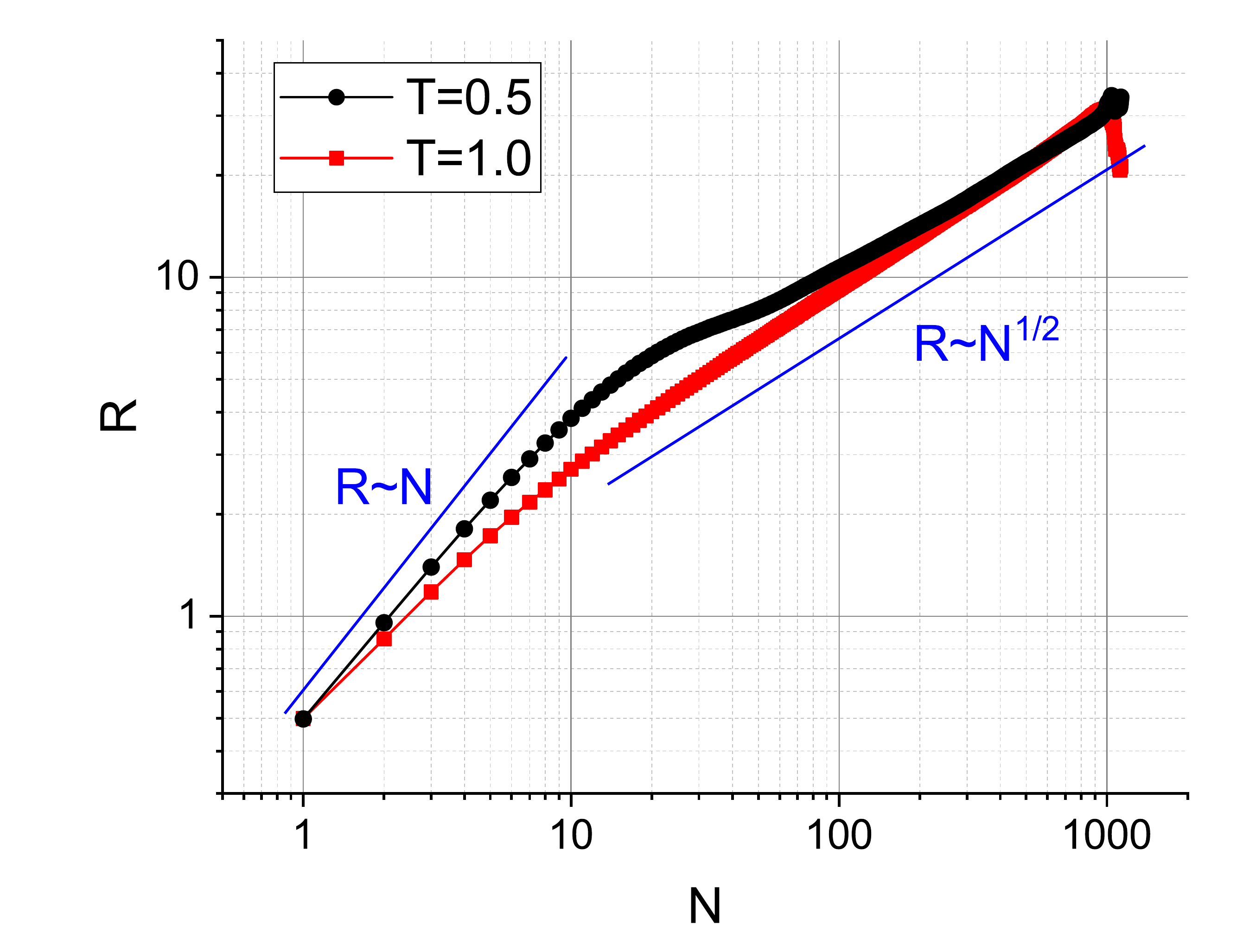}
\caption{R(S) for systems of 184000 particles at $T=0.4$ and $T=1.0$.}
\label{fig:rs}
\end{figure}

Experimental studies consider partial crystallization of growing chains as a main factor to synthesize low-entangled melts. Therefore, in our simulations we needed to include the effect of chain crystallization. To do so, we performed simulations of chain crystallization from equilibrium melt to determine conditions necessary for crystallization. Equilibrium melt was synthesized according to the following procedure. Polymerization with reaction rate $r=10^{-2}$ was carried out in the two systems ("smaller" and "larger", Fig. \ref{crystallization}) with fixed fraction of initiators $c=10^{-3}$. Temperature was set $T=1.0$, cutoff radius for all interactions was set $R_{cut}=1.02\sigma$. After complete polymerization (Fig. \ref{conversion}) we performed cooling of the equilibrium melt with constant rate $10^{-6}\tau^{-1}$, as described by \cite{luo2009coding}, to the final temperature $T=0.4$. Heating was performed to determine hysteresis of the crystallization-melting process \cite{luo2009coding,meyer2002formation}. Therefore, we determined, that conditions, described in the main text, section Molecular Dynamics Model, are sufficient for melt crystallization. In particular, we determined the temperature $T=0.5$, used in all our simulations of polymerization with simultaneous crystallization. Additionally, we determined the dependency of the average spatial distance between beads on the distance along the chain, $R(s)$ (Fig. \ref{fig:rs}). 

\subsection{Collapse of a single chain in a poor solvent}
As we have described in the previous section, we decided to take the temperature $T=0.5$ to simulate the crystallization in the melt. However, we should also adjust the interaction potential to model the poor solvent conditions to be able to simulate the growing chains, which simultaneously crystallize.

There are two ways to simulate the poor solvent conditions: to increase attraction of the polymer beads or to increase attraction of the solvent beads. We chose to follow the second way and reduced the cutoff radius of the LJ96 potential for the solvent-solvent interactions. Since $R_{cut}=1.02\sigma$ is the minimum of the LJ96 potential with the selected coefficients \cite{luo2009coding}, making the solvent-solvent interactions cutoff radius $R_{cutSS}$ smaller than $R_{cut}$ effectively reduces the effective size of the solvent beads. Therefore, repulsion between the solvent beads decreases and the solvent becomes poor to the polymer. We consequently reduced the $R_{cutSS}$ to find the coil-globule transition of the single chain length $N=1250$ beads, $T=0.5$. Collapse occurred with the value $R_{cutSS}=0.93\sigma$ after $\approx 10^{7}$ MD steps (or $\approx 10^{5}$ time units). To ensure fast collapse of the long chains, we chose the value $R_{cutSS}=0.9\sigma$.

\begin{figure}[h!]
    \centering
	\begin{subfigure}{0.2\textwidth}
	\includegraphics[width=\linewidth,height=\textheight,keepaspectratio]{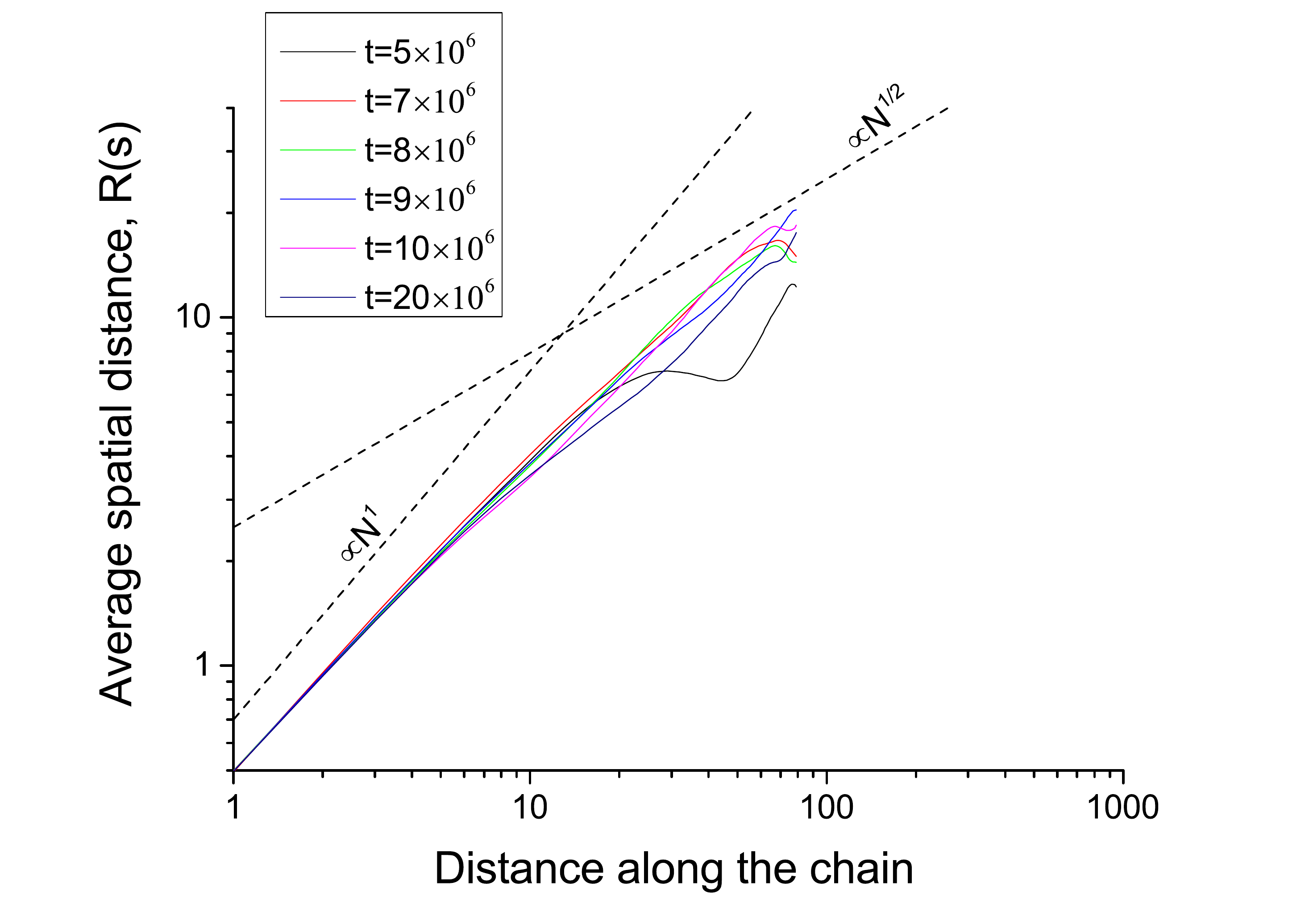}
	\subcaption{}
	\end{subfigure}
	\begin{subfigure}{0.2\textwidth}
	\includegraphics[width=\linewidth,height=\textheight,keepaspectratio]{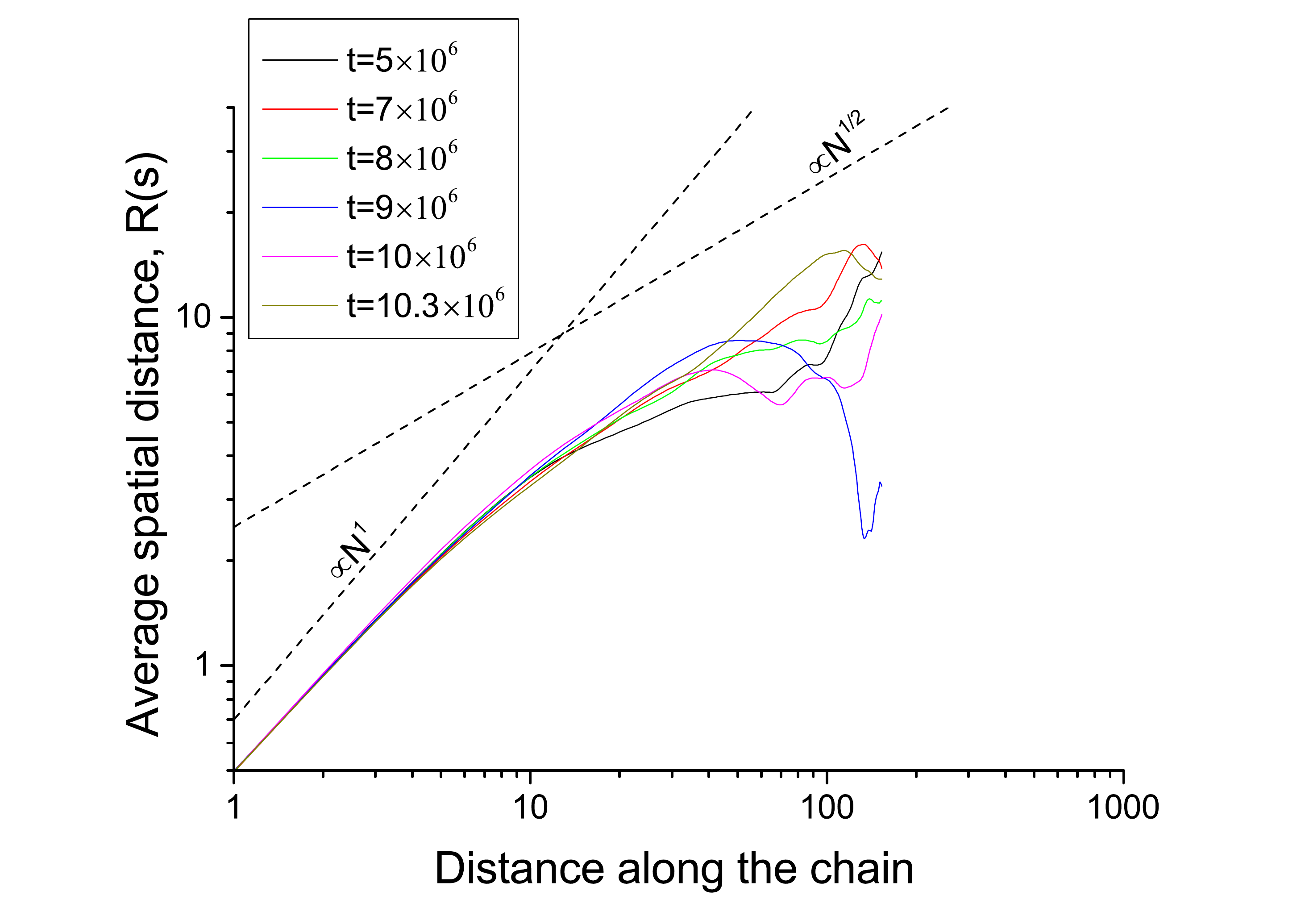}
	\subcaption{}
	\end{subfigure}
	\begin{subfigure}{0.2\textwidth}
	\includegraphics[width=\linewidth,height=\textheight,keepaspectratio]{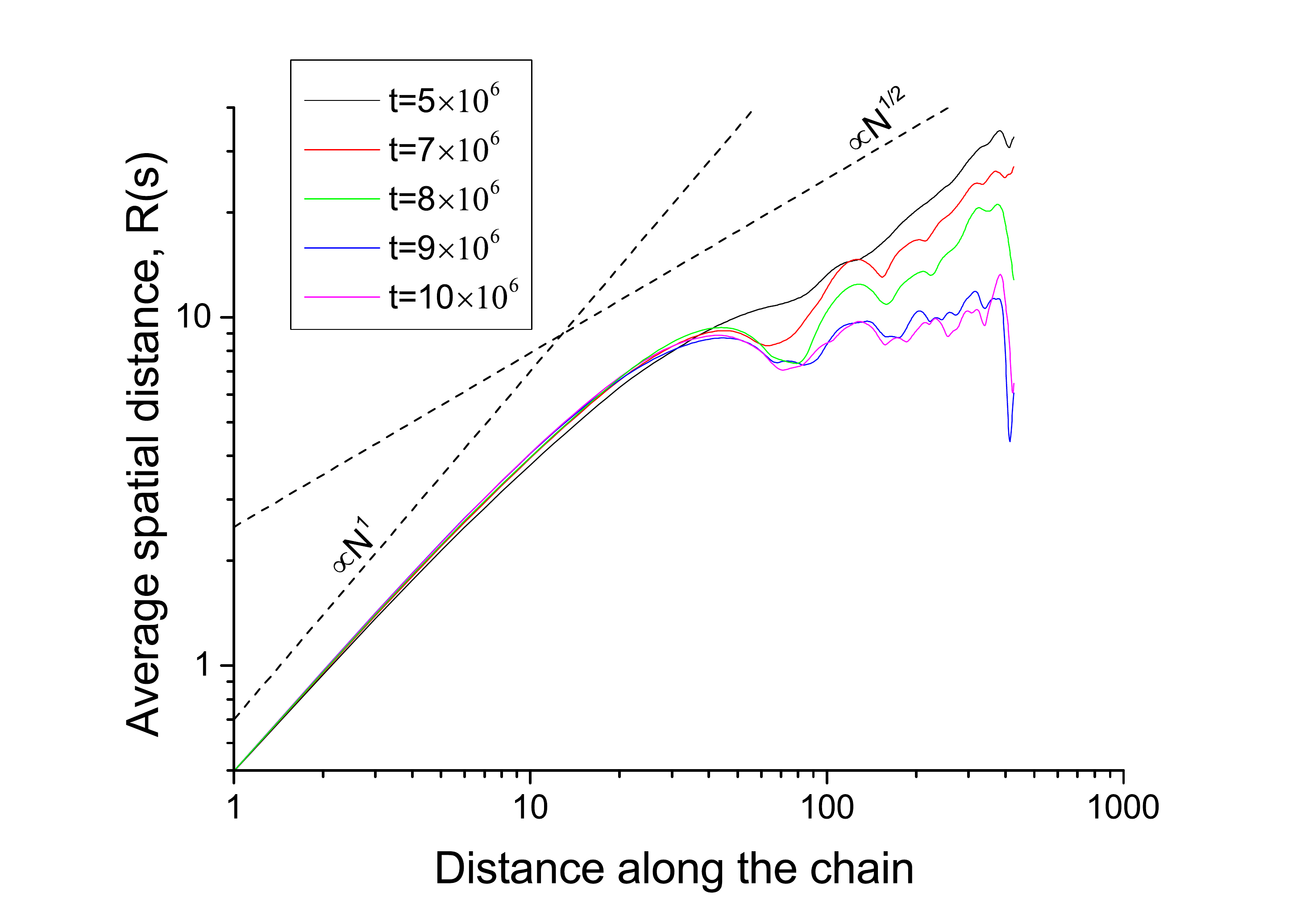}
	\subcaption{}
	\end{subfigure}
	\begin{subfigure}{0.2\textwidth}
	\includegraphics[width=\linewidth,height=\textheight,keepaspectratio]{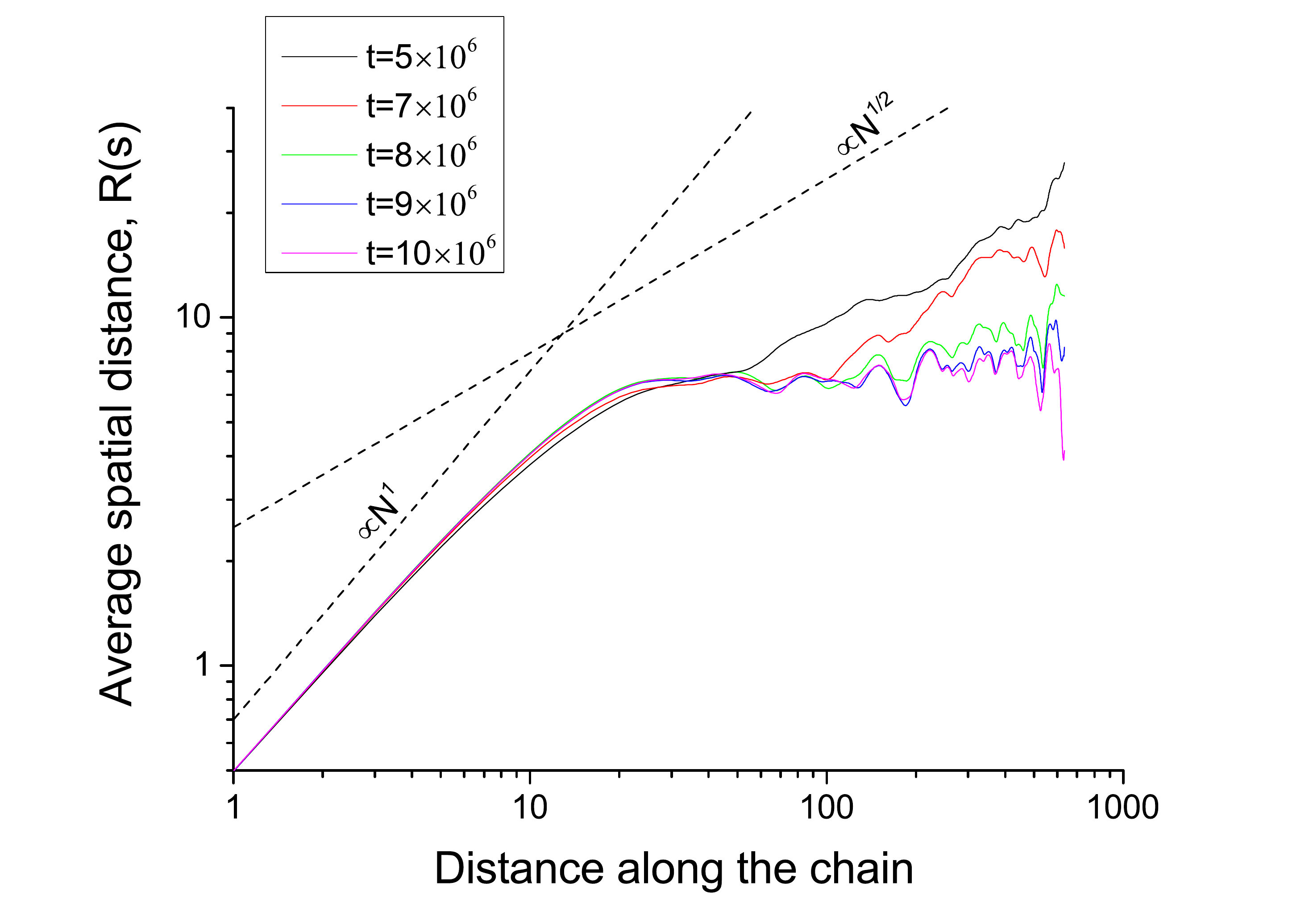}
	\subcaption{}
	\end{subfigure}
    \caption{$R(s)$ dependencies during collapse for different chain lengths: a) $N=79$, b) $N=153$, c) $N=426$ and d) $N=623$ beads. Time after collapse started is listed in the legend in MD time steps.}
    \label{coilglob}
\end{figure}

To characterize the effect of chain length on collapse with $R_{cutSS}=0.9\sigma$ and $T=0.5$, we simulated collapse of several chains length $\propto 10^2$ beads (Fig. \ref{coilglob}). Initial structures for the simulations of collapse were prepared according to the following procedure. Process of chain growth with reaction rate $0.01$ was carried out in the $T=0.5$ in the box size $40\sigma$ with periodic boundary conditions (PBC), and with $54.5\times10^3$ particles and 1 initiator. Cutoff radius for all interactions via LJ96 was set $R_{cut}=1.02\sigma$. During this process systems with chains of different lengths in the swollen coil state were obtained. Then the $R_{cutSS}$ was set to $0.9\sigma$ and the kinetics of collapse was studied, starting from the swollen coil state. To avoid interactions of the "ghost" beads with "real" beads via PBC during collapse, we studied collapse of rather short chains: $N=79$, $N=153$, $N=426$ and $N=623$ beads (Fig. \ref{coilglob}). 

We observed, that chain collapse behavior depends in a very non-trivial way on the chain length. First, if the chain length is $N=79$, it does not collapse, and resembles a swollen coil (Fig. \ref{coilglob}a). If we double the chain length and see on the behavior of the $N=153$ beads case (Fig. \ref{coilglob}b), we observe the following picture. After $t=9\times10^6$ MD time steps the chain resembles a transient state at the beginning of collapse, but then fluctuates back to the swollen state after $\approx 10.3\times10^6$ MD time steps. If we increase the chain length to the $N=426$ beads, we observe a well-defined collapse after $t\approx 9\times10^6$ MD time steps (Fig. \ref{coilglob}c). The same behavior is observed in the case $N=633$ beads (Fig. \ref{coilglob}d).

Hence, we can say, that the chains length less than some value around $\approx100$ beads can not collapse to the crystallized globule with the given set of parameters. This fact is used for construction of the the high fraction of initiators regime in the analytical theory (see the main text).

\subsection{Growth of a single chain in a poor solvent}
\begin{figure}[h!]
\centering
\begin{subfigure}{0.2\textwidth}
	\includegraphics[width=\linewidth,height=\textheight,keepaspectratio]{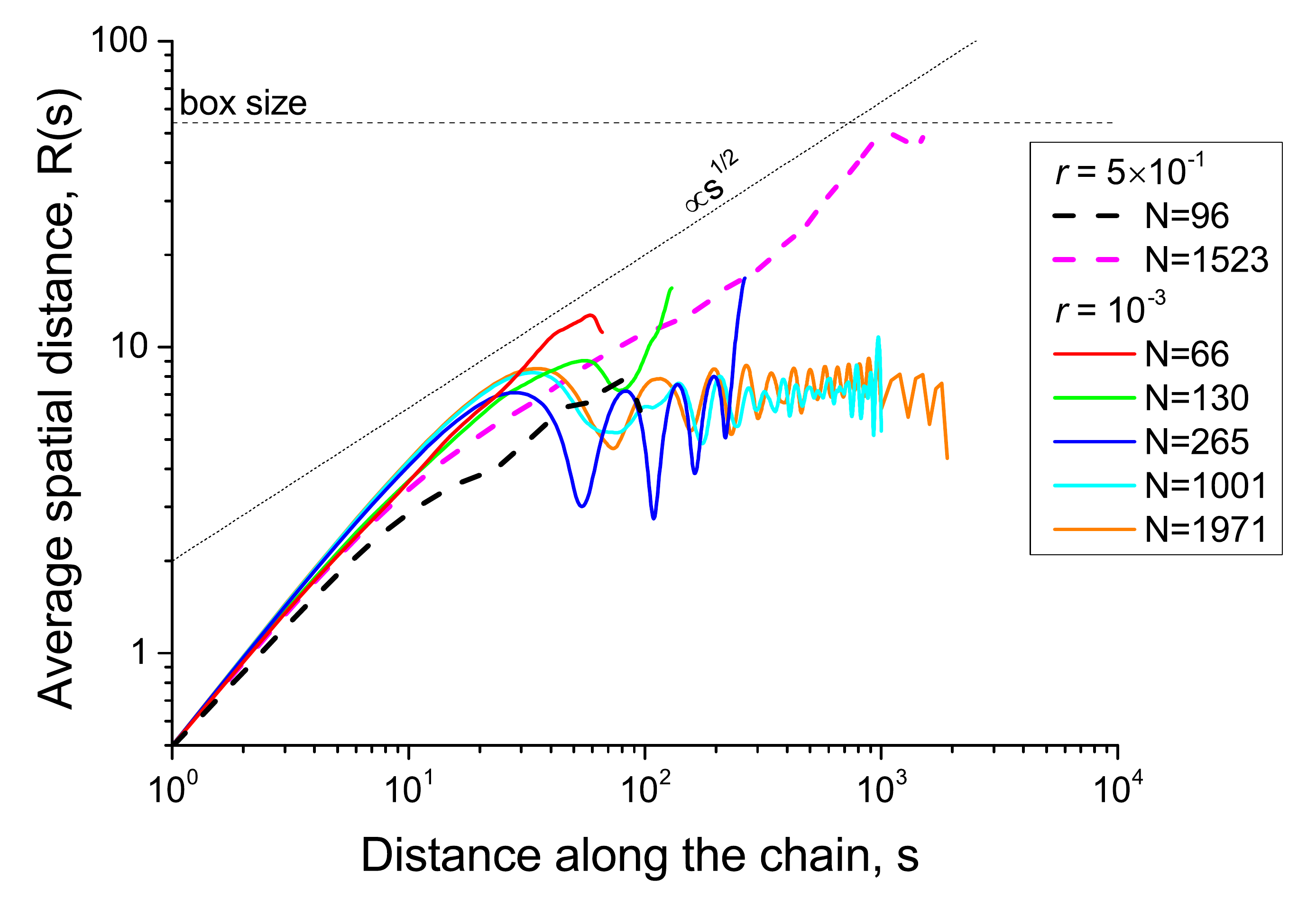}
	\subcaption{}
	\end{subfigure}
	\begin{subfigure}{0.2\textwidth}
	\includegraphics[width=\linewidth,height=\textheight,keepaspectratio]{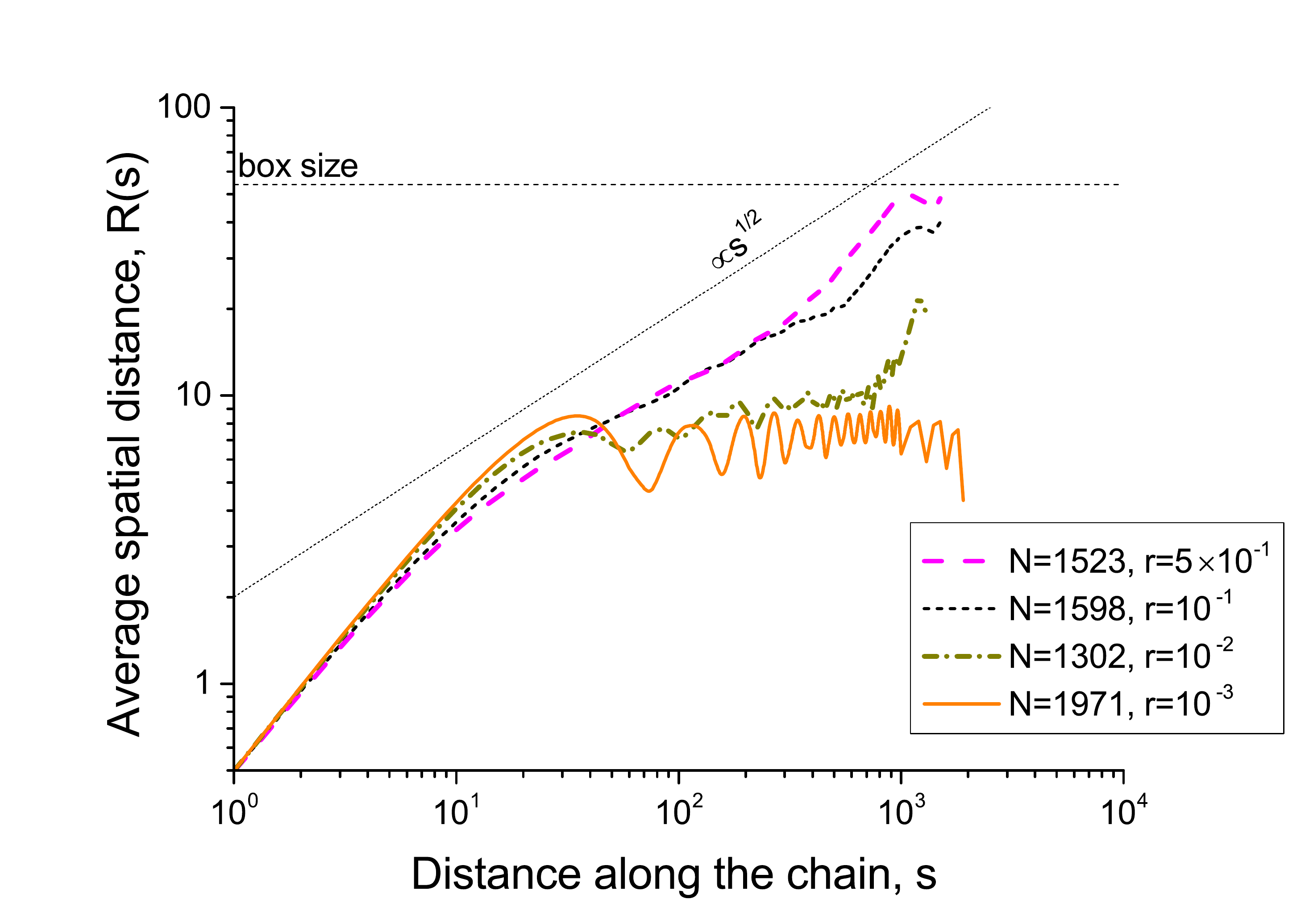}
	\subcaption{}
	\end{subfigure}
\caption{$R(s)$ dependencies for a single chain growing in a poor solvent with different reaction rate $r$. The values of $r$ are listed in the legend in reverse time units. a) Comparison of the growth processes with the smallest and the largest reaction rates in our simulations. b) Comparison of long chain conformations synthesized with 4 different reaction rates.}
\label{singlechaingrowth}
\end{figure}

To study the effect of the reaction rate on the growing chain conformation, we performed simulations of a single chain growth. We analyzed conformations using the same method as in the previous section, i.e. $R(s)$ dependencies (Fig. \ref{singlechaingrowth}). As in the previous section, we stopped simulations at some moment to prevent interactions of the chain with itself via PBC. We used a larger system to study longer chains: $184\times 10^3$ particles and 1 initiator. Temperature conditions and LJ96 coefficients were set as in the simulations of the many growing chains (see Molecular Dynamics Model section in the main text).

We discovered, that the reaction rate $r$ strongly affects the conformation of a long growing chain. As we determined in the previous section, short enough chains can not collapse in the chosen conditions, so the conformations of short chains do not depend on the reaction rate and remain non-collapsed (except for the slight deviations in the persistence length, that remains $\approx10$ beads, Fig. \ref{singlechaingrowth}a). However, if the chain length exceeds $\approx 130$ beads, the chain being slowly synthesized collapses, and the fast growing chain remains Gaussian (Fig. \ref{singlechaingrowth}a). Snapshot of the slowly growing chain ($r=10^{-3}$) is presented in the main text, Fig. 1c.

We also studied, how conformation of a long chain depends on the reaction rate in the intermediate range of $r$ (Fig. \ref{singlechaingrowth}b). We observed, that decrease of polymerization rate led to compactification of growing long chains. Chain growing with $r=10^{-1}$ is still swollen on the large scale, but is compactified on the local scale (see snapshot in the main text, Fig. 1d). This supports the "blob" picture, suggested in the analytical model of the fast growing long chains. Decrease of the reaction rate to the $r=10^{-2}$ leads to the transient structure, which does not resemble a single globule, but is not swollen Gaussian chain on the global scale (Fig. \ref{singlechaingrowth}b).

\subsection{Polymerization process in various systems: analysis of conformations}
In our analytical theory we described the process of chain growth, and, in particular, entanglement of chains and melt formation. We have mentioned, that the exponent $k$, which determines the power law decrease of $N_e(N)$ dependency in the second phase, may depend on $r$ or $c$, and, in fact, describes the structure of entangled melt. In this section we describe the average chain conformation in different systems studied in our work by MD simulation.

\begin{figure}[h!]
\centering
	\begin{subfigure}{0.2\textwidth}
	\includegraphics[width=\linewidth,height=\textheight,keepaspectratio]{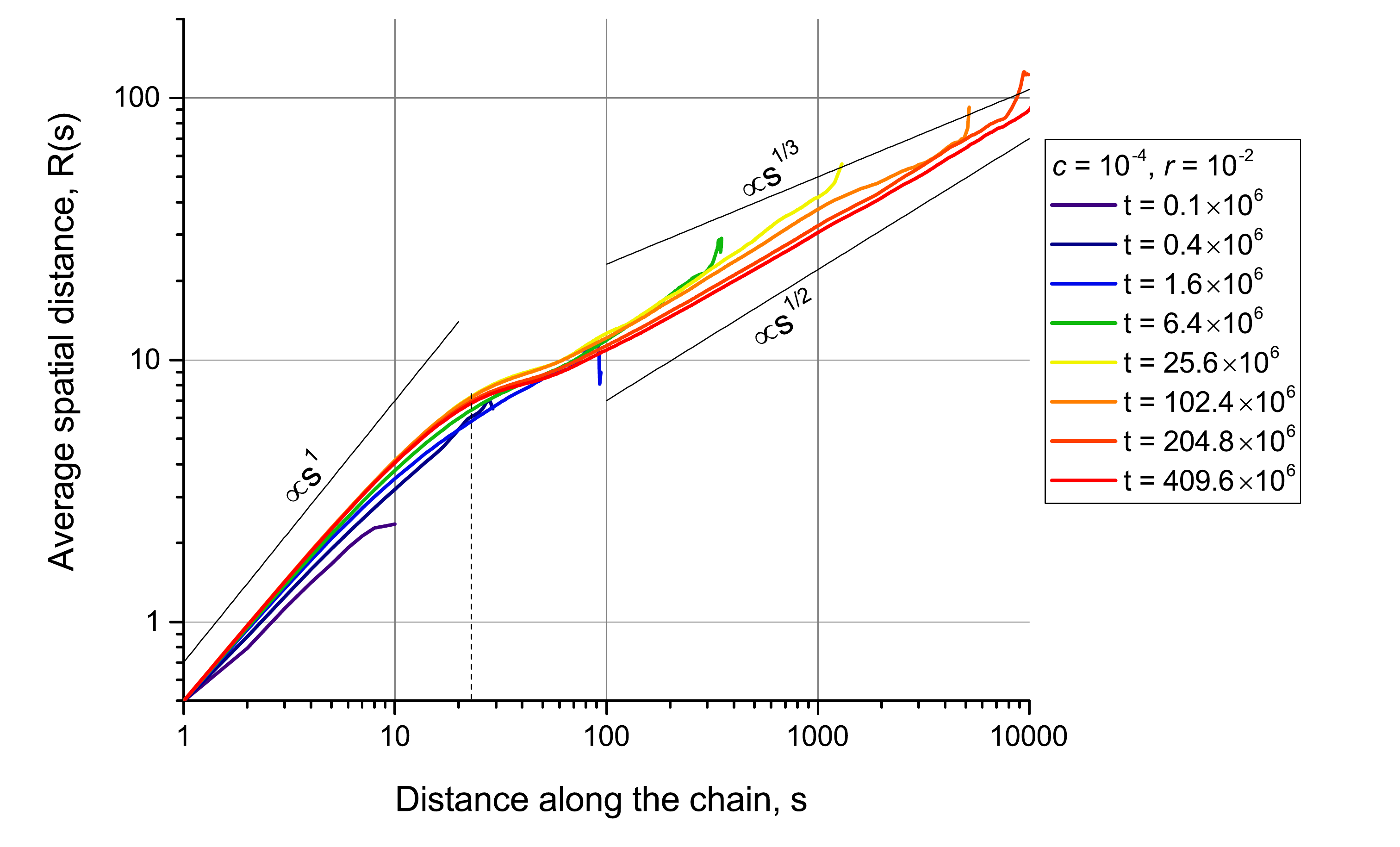}
	\subcaption{}
	\end{subfigure}
	\begin{subfigure}{0.2\textwidth}
	\includegraphics[width=\linewidth,height=\textheight,keepaspectratio]{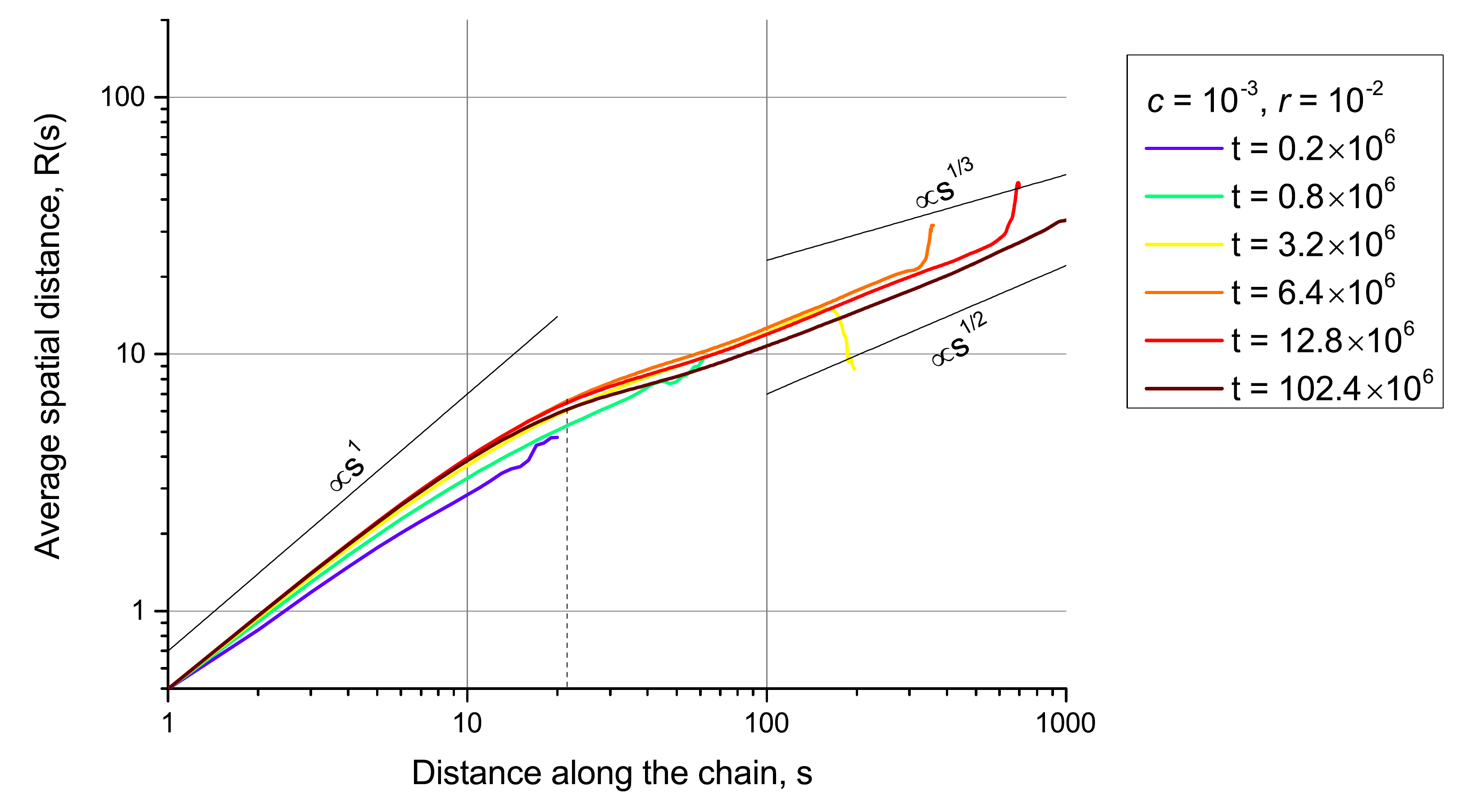}
	\subcaption{}
	\end{subfigure}
	\begin{subfigure}{0.2\textwidth}
	\includegraphics[width=\linewidth,height=\textheight,keepaspectratio]{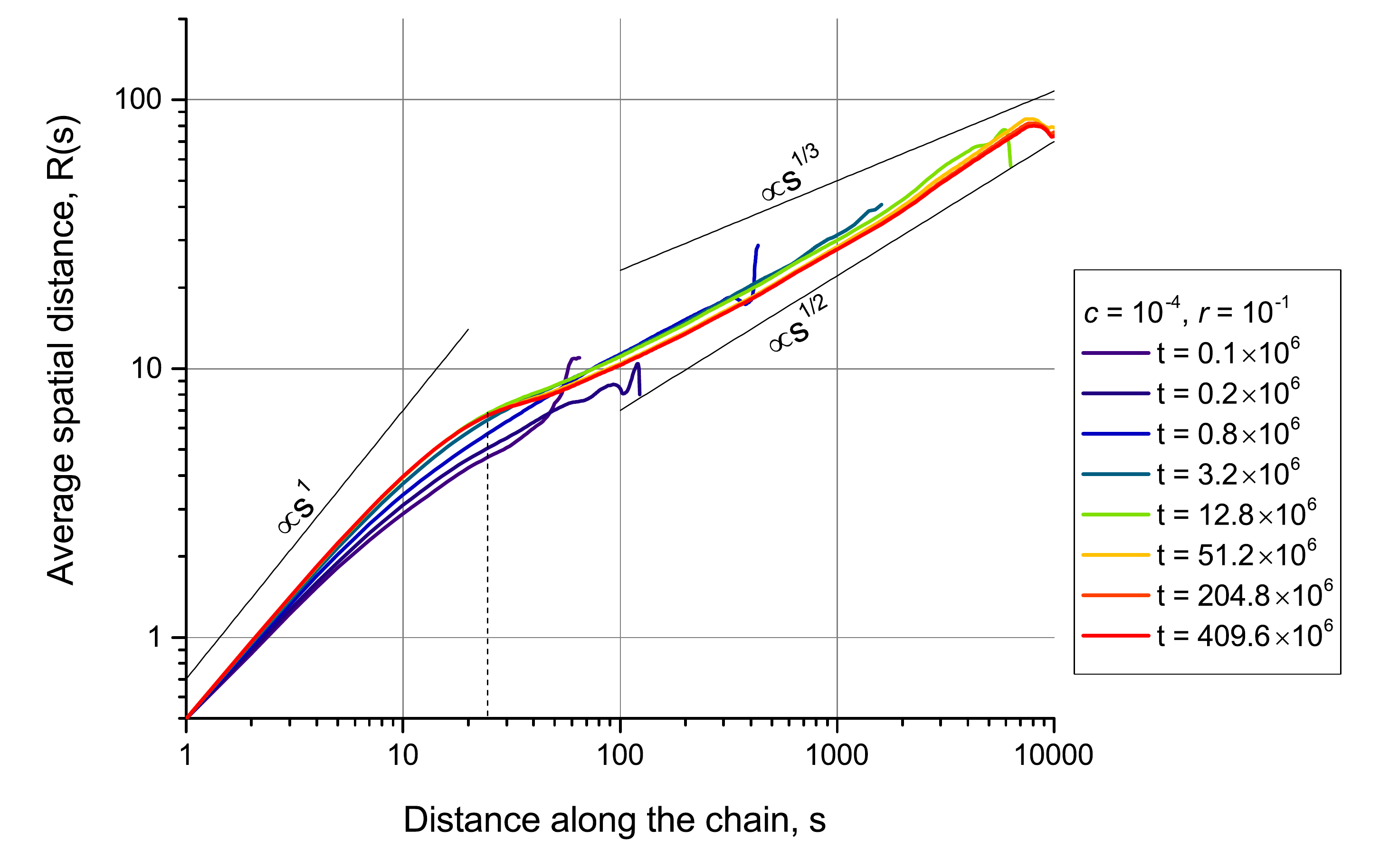}
	\subcaption{}
	\end{subfigure}
	\begin{subfigure}{0.2\textwidth}
	\includegraphics[width=\linewidth,height=\textheight,keepaspectratio]{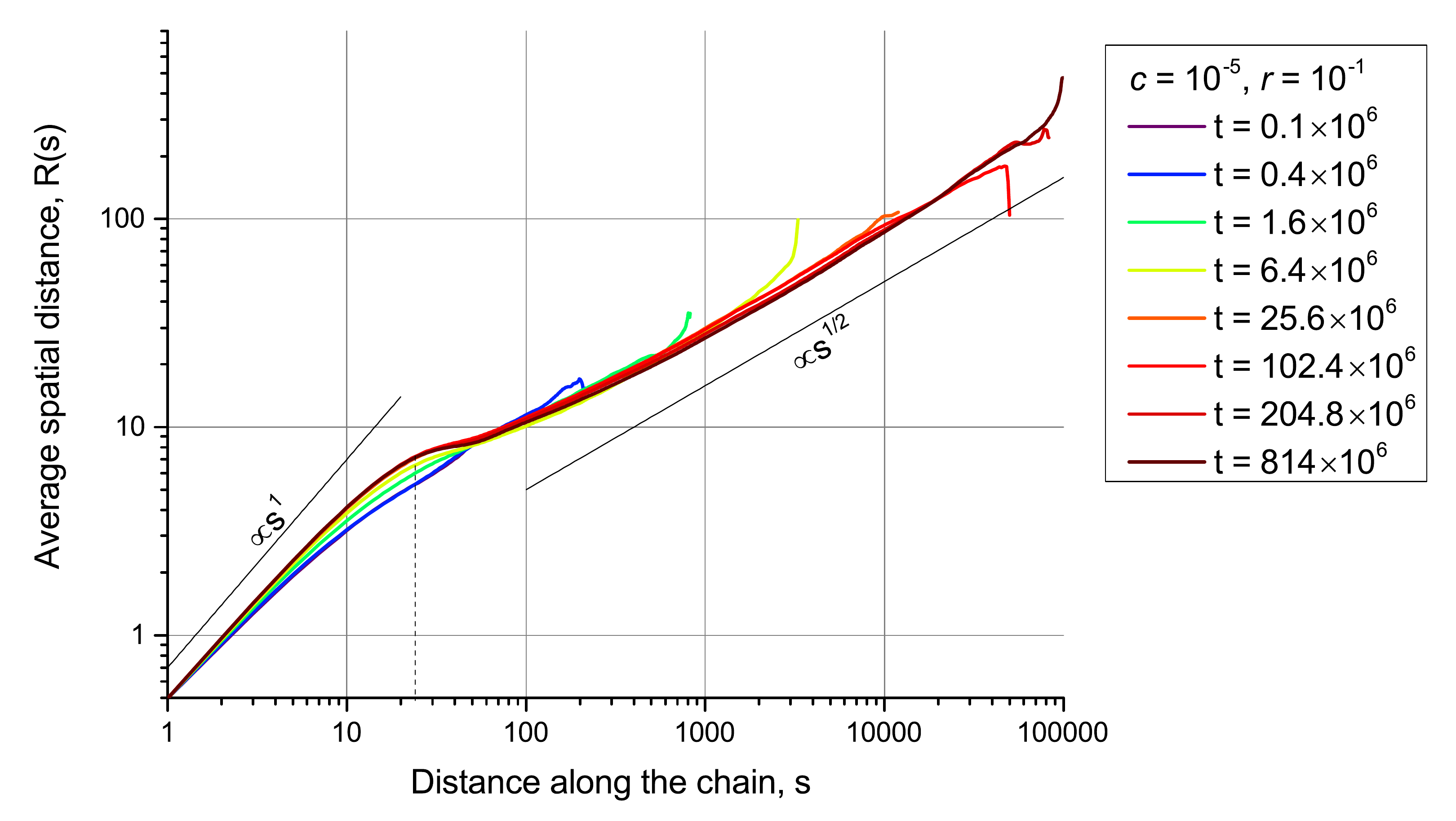}
	\subcaption{}
	\end{subfigure}
	\begin{subfigure}{0.2\textwidth}
	\includegraphics[width=\linewidth,height=\textheight,keepaspectratio]{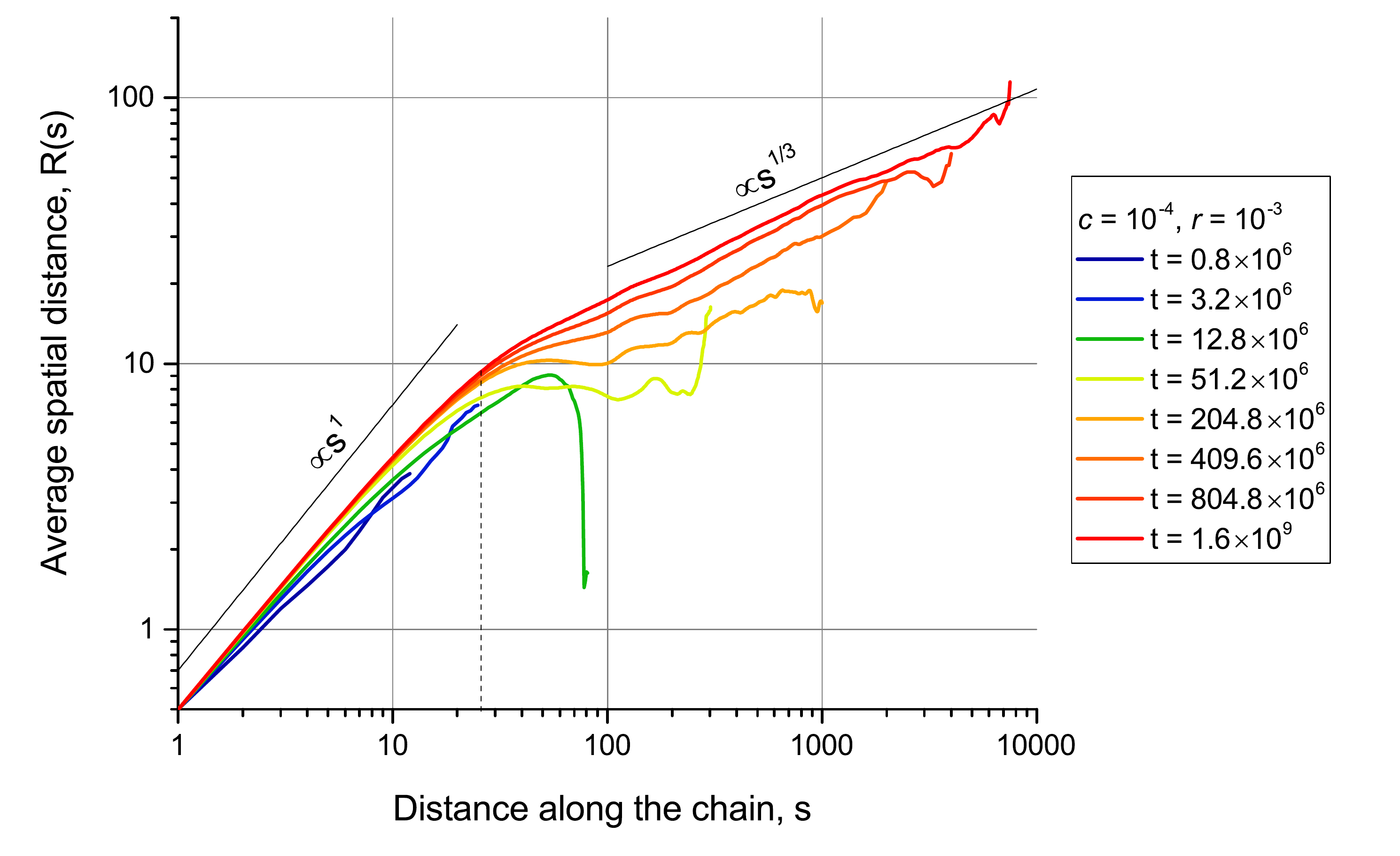}
	\subcaption{}
	\end{subfigure}
\caption{$R(s)$ dependencies for the systems of many growing chains. a), b) - reaction rate $r=10^{-2}$, fraction of initiators $c=10^{-4}$ and $c=10^{-3}$, respectively. c), d) - reaction rate $r=10^{-1}$, fraction of initiators $c=10^{-4}$ and $c=10^{-5}$, respectively. e) reaction rate $r=10^{-3}$, fraction of initiators $c=10^{-4}$. Black solid lines represent $R(s)$ scaling on different length scales.}
\label{rs}
\end{figure}

\begin{figure}[h!]
\centering
	\begin{subfigure}{0.2\textwidth}
	\includegraphics[width=\linewidth,height=\textheight,keepaspectratio]{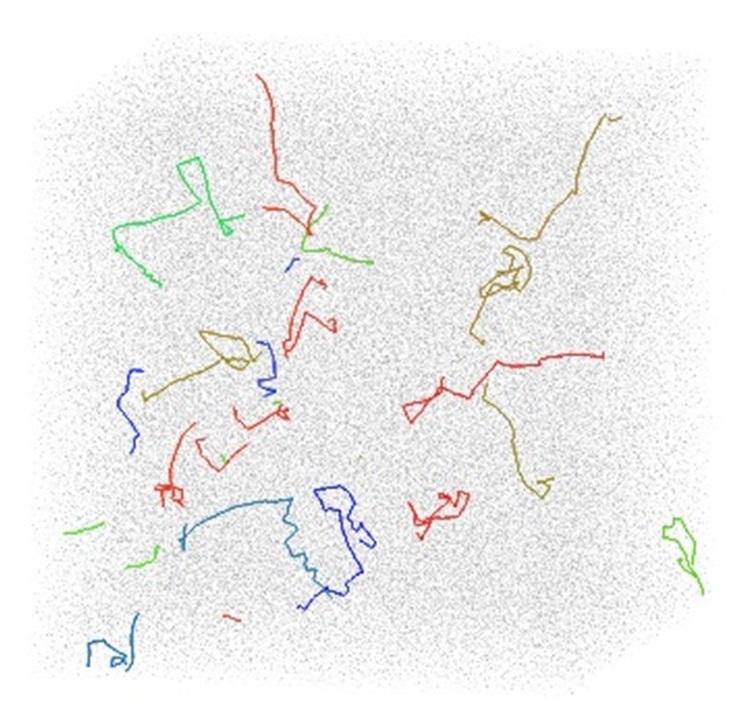}
	\subcaption{}
	\end{subfigure}
	\begin{subfigure}{0.2\textwidth}
	\includegraphics[width=\linewidth,height=\textheight,keepaspectratio]{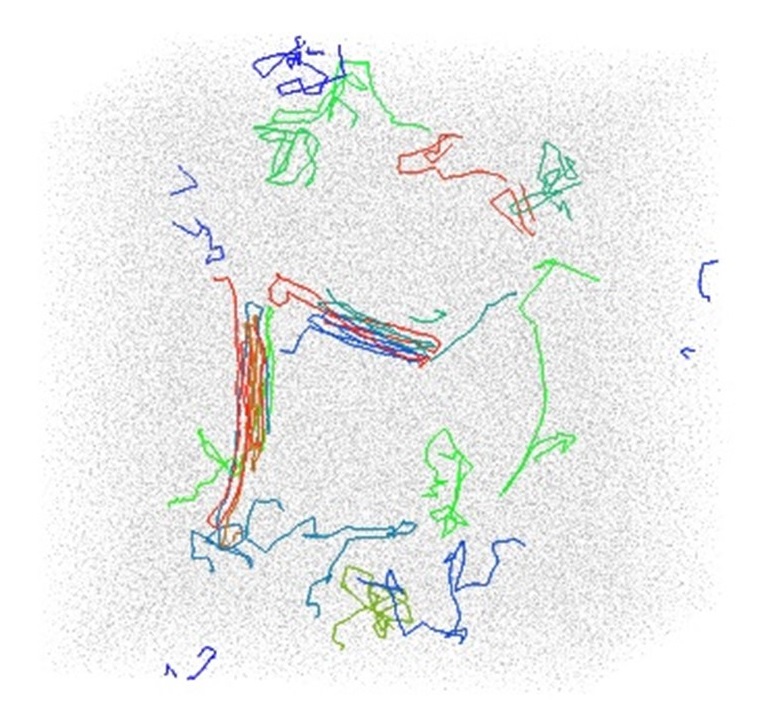}
	\subcaption{}
	\end{subfigure}
	\begin{subfigure}{0.2\textwidth}
	\includegraphics[width=\linewidth,height=\textheight,keepaspectratio]{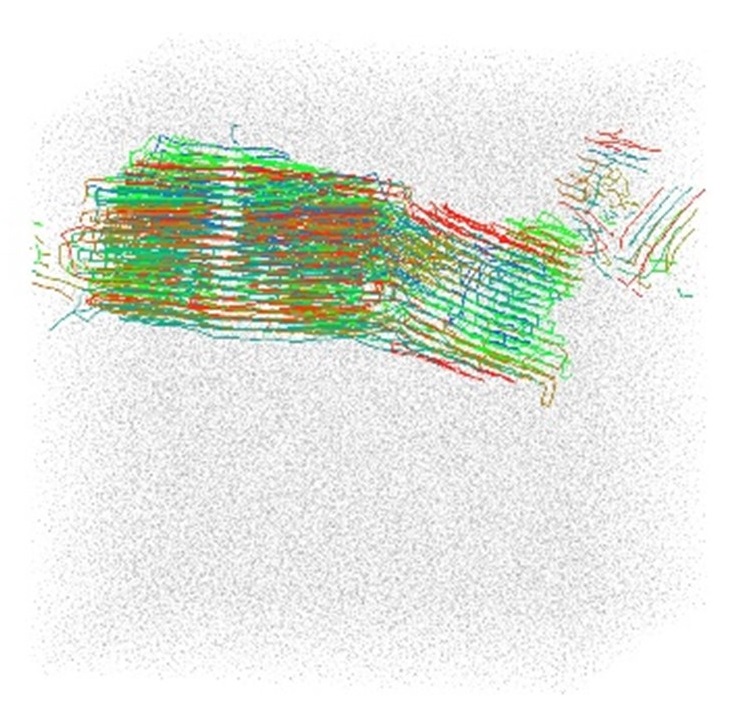}
	\subcaption{}
	\end{subfigure}
	\begin{subfigure}{0.2\textwidth}
	\includegraphics[width=\linewidth,height=\textheight,keepaspectratio]{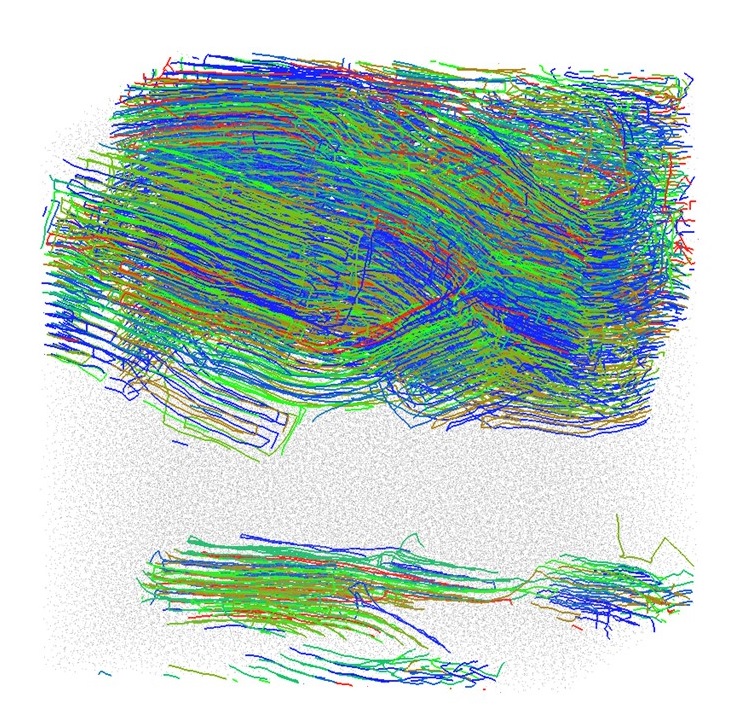}
	\subcaption{}
	\end{subfigure}
\caption{Snapshots taken during chain growth in the system with $c=10^{-4}$ and $r=10^{-3}$.}
\label{snapsslow}
\end{figure}

First we studied the influence of the reaction rate $r$ on the average chain conformation. We compared $R(s)$ scaling in the three systems with $r=10^{-3}$, $r=10^{-2}$ and $r=10^{-1}$, fraction of initiators $c=10^{-4}$ (Fig. \ref{rs}a, \ref{rs}c, \ref{rs}e). We see, that $r$ affects the average chain conformation during melt synthesis. The $R(s)$ exponent tends to the value $1/3$ for slow reactions, and to the value $1/2$ for fast reactions. In other words, chains are more compact on the large length scale in the melt obtained after slow synthesis, than Gaussian conformation, observed in the melt crystallized from the equilibrium melt (Fig. \ref{fig:rs}), and in the melt obtained after fast polymerization (Fig. \ref{rs}c).

We have also studied the effect that the fraction of initiators has on the average chain conformation (Fig. \ref{rs}a, \ref{rs}b, \ref{rs}c, \ref{rs}d). In both pairs of systems (with $r=10^{-1}$ and $r=10^{-2}$) we have not observed significant difference in $R(s)$ scaling exponent.

\subsection{Effect of the system size on the average entanglement length}
\begin{figure}[h!]
\centering
	\begin{subfigure}{0.2\textwidth}
	\includegraphics[width=\linewidth,height=\textheight,keepaspectratio]{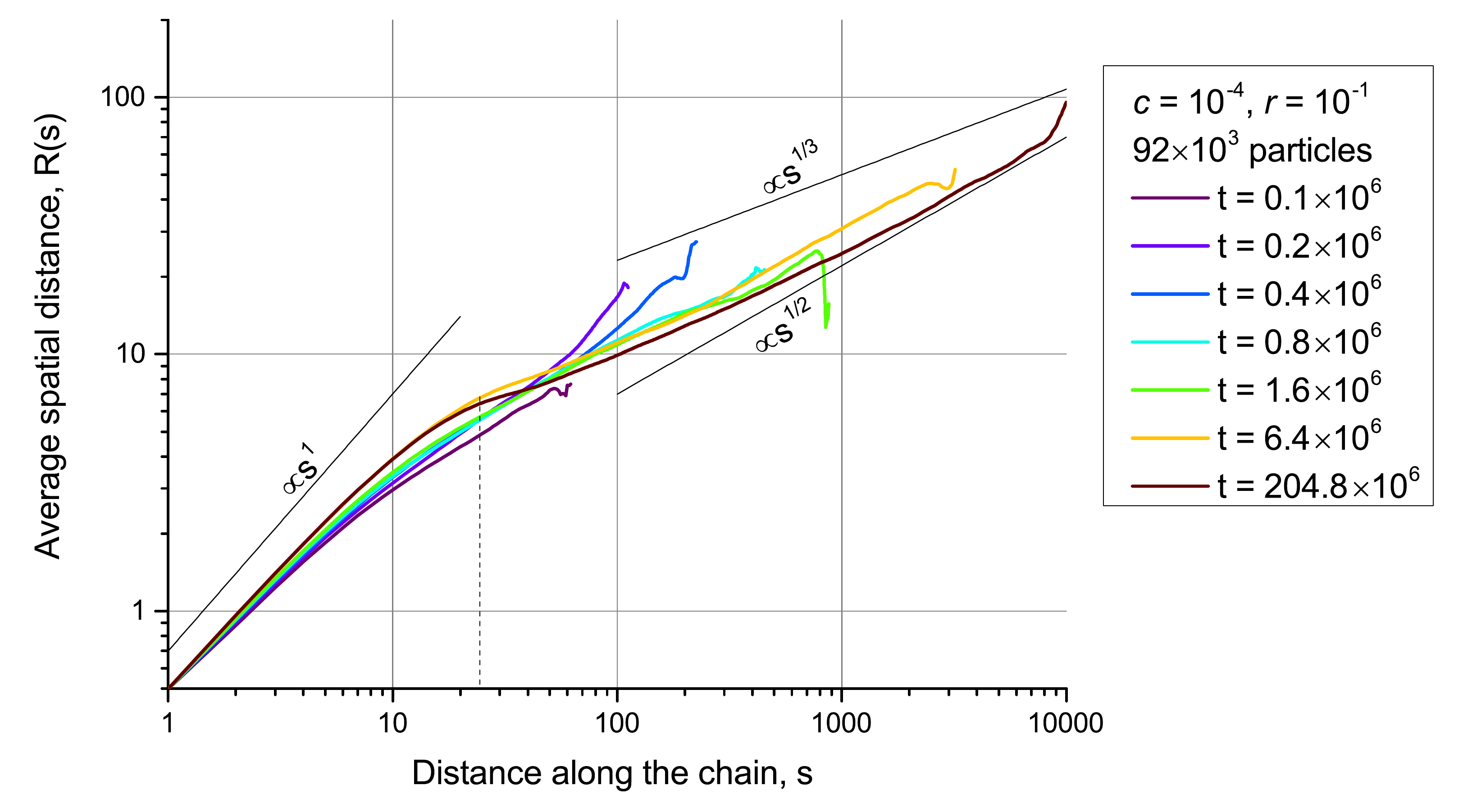}
	\subcaption{}
	\end{subfigure}
	\begin{subfigure}{0.2\textwidth}
	\includegraphics[width=\linewidth,height=\textheight,keepaspectratio]{rs1.pdf}
	\subcaption{}
	\end{subfigure}
\caption{$R(s)$ dependencies for the systems of many growing chains, reaction rate $r=10^{-1}$, fraction of initiators $c=10^{-4}$. a) $92\times10^3$ particles. b) $184\times10^3$ particles. Black solid lines represent $R(s)$ scaling on different length scales.}
\label{sizeeffect}
\end{figure}

\begin{figure}[h!]
\centering
  \includegraphics[width=0.5\linewidth]{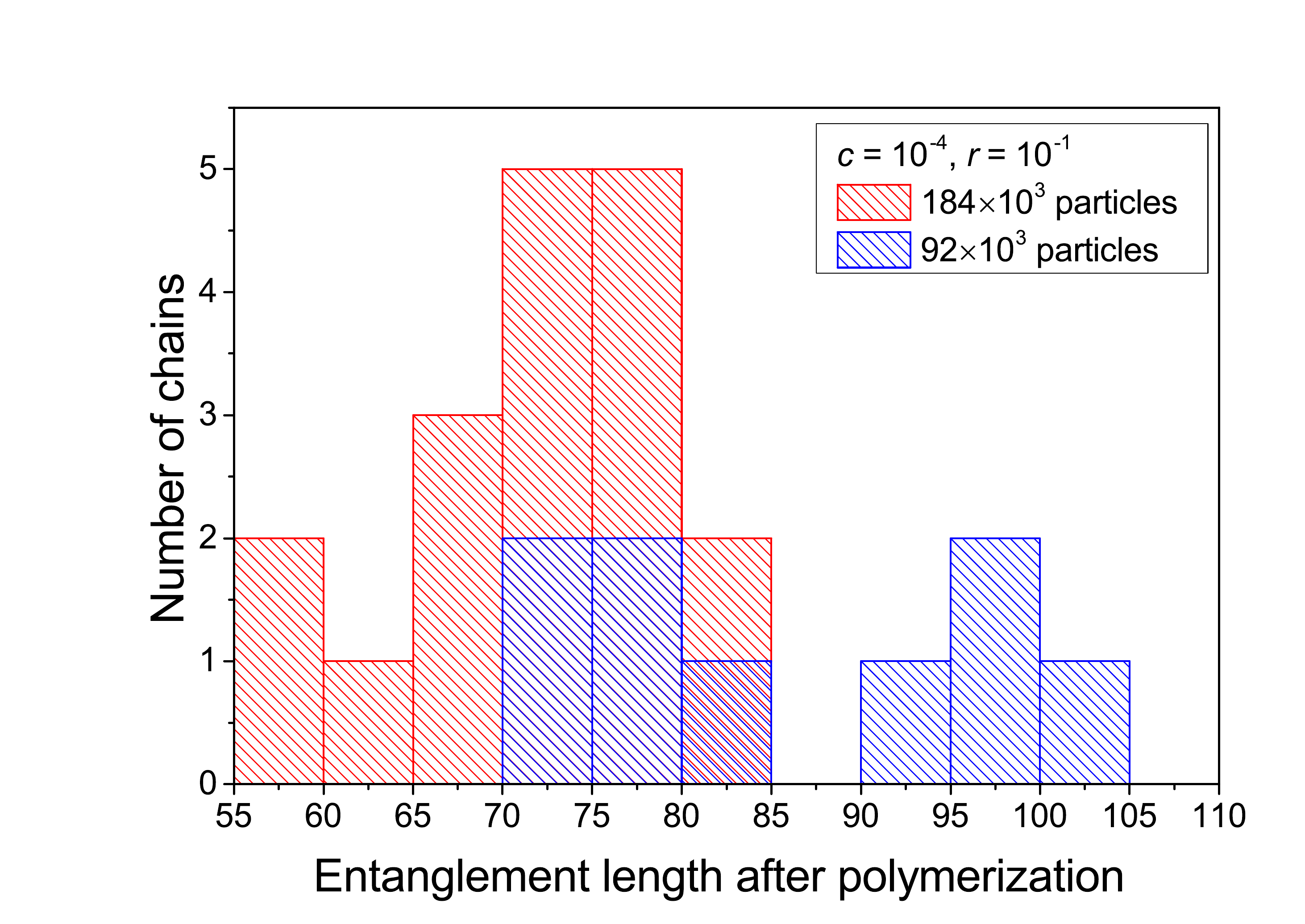}
\caption{Distributions of $N_e(N\approx1/c)$ across different chains in the two systems with $c=10^{-4}$ and $r=10^{-1}$, but with different number of particles (blue and red histogram represents the smaller and the larger system, respectively.}
\label{nedistrcomp}
\end{figure}

As we have described in the Computer Simulation Results section in the main text, systems with different fraction of initiators $c$ contain different number of initiators (184, 18 and 9 initiators in the systems with $c=10^{-3}$, $c=10^{-4}$ and $c=10^{-5}$, respectively). Therefore, the effect of $N_e(N=1/c)$ increase after decrease of $c$ (Fig. 5a in the main text) may be due to decrease of the number of initiators (i.e. size-effect). To exclude this factor, we performed simulations of the system with $c=10^{-4}$ and $r=10^{-1}$, but with 9 initiators and $92\times 10^{3}$ particles (i.e. this system is twice smaller, than the system discussed in the main text). First we analyzed the $R(s)$ dependencies during chain growth in both cases (Fig. \ref{sizeeffect}). We observed, that the average chain conformation does not change during melt synthesis if the system is twice smaller. Secondly, we built distributions of $N_e(N\approx1/c)$ in both systems (Fig. \ref{nedistrcomp}). We observed, that the distributions are strongly intersecting and the values of $N_e(N\approx1/c)$ coincide within standard deviation: $71\pm8$ in the larger system and $85\pm11$ in the smaller system. Therefore, we conclude, that size of the system does not affect dependencies measured in the work: $N_e(N)$ and $R(s)$.

\section{References}
\bibliography{sample}
\end{document}